\documentclass[10pt,journal,compsoc]{IEEEtran}

\ifCLASSOPTIONcompsoc
  \usepackage[nocompress]{cite}
\else
  \usepackage{cite}
\fi

\ifCLASSINFOpdf
\else
\fi

\usepackage{booktabs}
\usepackage{dcolumn}
\usepackage[british]{babel}
\usepackage{siunitx}
\usepackage{amsfonts}
\usepackage{subfigure}
\usepackage{overpic}
\usepackage{color}

\usepackage{multirow}

\renewcommand{\vec}[1]{{\mathbf{#1}}} 
\newcommand{\Dt}{\Delta t}
\newcommand{\particle}{p}
\newcommand{\vecp}{{\vec{x}_p}}

\newcommand{\vecx}{{\vec{x}}}
\makeatletter
\newcommand*{\rom}[1]{\expandafter\@slowromancap\romannumeral #1@}
\makeatother

\newcommand{\oil}{}
\newcommand{\dropcoll}{}
\newcommand{\jet}{}
\def \oil/{\textit{Oil Inclusions}}
\def \dropcoll/{\textit{Drop Collision}}
\def \jet/{\textit{Non-Newtonian Jet}}

\usepackage{algorithm}
\usepackage{algorithmicx}
\usepackage{algpseudocode}

\algnewcommand\algorithmicforeach{\textbf{for each}}
\algdef{S}[FOR]{ForEach}[1]{\algorithmicforeach\ #1\ \algorithmicdo}
\floatname{algorithm}{Algorithm}

\usepackage{color}
\definecolor{grey}{rgb}{0.4, 0.4, 0.4}
\definecolor{darkgreen}{rgb}{0.0, 0.6, 0.0}

\begin{document}

\title{Visualization of Feature Separation in Advected Scalar Fields}

\author{Grzegorz K. Karch, Filip Sadlo, Sebastian Boblest, \\ Moritz Ertl, Bernhard Weigand, Kelly Gaither, Thomas Ertl
}

\markboth{Journal of \LaTeX\ Class Files,~Vol.~14, No.~8, August~2015}%
{Shell \MakeLowercase{\textit{et al.}}: Bare Advanced Demo of IEEEtran.cls for IEEE Computer Society Journals}

\IEEEtitleabstractindextext{%
\begin{abstract}
Scalar features in time-dependent fluid flow are traditionally visualized
using 
3D representation, and their topology changes over time are often conveyed with abstract graphs.
Using such techniques, however, the structural details of feature separation and the temporal evolution of features undergoing topological changes are difficult to analyze.
In this paper, we propose a novel approach for the spatio-temporal visualization of feature separation that segments feature volumes into regions with respect to their contribution to distinct features after separation.
To this end, we employ particle-based feature tracking to find volumetric correspondences between features at two different instants of time.
We visualize this segmentation by constructing mesh boundaries around each volume segment of a feature at the initial time that correspond to the separated features at the later time. To convey temporal evolution of the partitioning within the investigated time interval, we complement our approach with spatio-temporal separation surfaces.
For the application of our approach to multiphase flow, we additionally present a feature-based corrector method to ensure phase-consistent particle trajectories.
The utility of our technique is demonstrated by application to 
two-phase (liquid-gas) and multi-component (liquid-liquid) flows where the scalar field represents the fraction of one of the phases.
\end{abstract}

}

\maketitle

\IEEEdisplaynontitleabstractindextext

%
\IEEEpeerreviewmaketitle

\ifCLASSOPTIONcompsoc
\IEEEraisesectionheading{\section{Introduction}\label{sec:introduction}}
\else
\section{Introduction}
\label{sec:introduction}
\fi

\IEEEPARstart{F}{low} visualization is widely used for the analysis of natural and engineering processes related to fluid motion.
As simulation data continuously grow in size, mainly due to increasing computational power of both supercomputers and consumer desktops, feature extraction and tracking has gained particular attention.
In feature-based visualization, visual data representation is reduced to important characteristics, typically physical quantities, flow topology, and quantities derived from generic scalar fields~\cite{CGF:CGF723}.
These features are then tracked in time to determine topological events, such as feature birth or split.
The topology can be intuitively conveyed with a graph, where edges correspond to the features and nodes to the events.
Such an approach reduces the amount of information to the essential part, relevant to the research at hand, and therefore allows for effective analysis of potentially enormous data.

While such feature-level visualization reveals the overall feature topology dynamics, detailed spatial information on volumetric partitioning of features is difficult to obtain with existing techniques.
Thus, we propose a visualization approach that can reveal the separation dynamics within features at some initial time with reference to a later target time.
To this end, we extract boundaries around regions within a feature that correspond to other features in the course of time, and additionally convey the temporal information of the separation by means of separation surfaces.

In the investigation of topology in general time-dependent flow, ridge extraction in the finite-time Lyapunov exponent (FTLE) is a standard method for the determination of regions that exhibit different flow behavior.
The FTLE measures the separation of neighboring trajectories, and ridges in the FTLE separate those regions.
With our visualization method, features representing physical quantities can be analyzed from a similar view point: given a feature at some point in time, we want to know how it will evolve over the course of time.
However, whereas the ridges in the FTLE expose maximum separation of trajectories after a given time interval, our boundaries reveal separation of \emph{quantities} after a time interval.

Our visualization technique expands upon traditional feature visualization in several ways.
First, it allows for static visualization of dynamic processes, and therefore reduces visual clutter.
Second, it combines advantages of standard feature tracking methods and FTLE-based methods, in that it allows for a detailed inspection of the separation of features.
Third, it is applicable to a broad class of multiphase flow, including two-phase flow (liquid in gas surrounding) and multi-component flow (liquid-liquid or multi-component liquid in gaseous surrounding).

An important type of flow simulation is two-phase flow.
Example cases are simulations of droplet collisions or liquid jets.
The break-up of liquid jets into droplets appears in many technical applications as well as in nature.
Examples range from a simple faucet over spray painting and spray drying in food processing to fuel injection in combustion engines.
Especially in cases where the disruptive forces are much stronger than the cohesive forces, called atomization, the jets break up into many small droplets shortly after injection.
Another important line of research is liquid-liquid flow, where immiscible fluids (for instance, oil and water) interact, for example, in emulsification processes.
Such flow has often considerably different characteristics compared to two-phase flow, e.g., due to high viscosity of both components.
Analysis of these processes, which occur on both very small temporal and spatial scales, is still very difficult.

Our visualization approach helps with the analysis of features in multiphase flow by visualizing the changing topology with respect to the feature at a selected time, and hence providing a static representation of the dynamics of features.
To summarize, the main contribution of our work is a spatio-temporal visualization of feature separation dynamics that reveals the feature volume distribution among developed features as well as temporal information of the separation. As a necessary prerequisite, we propose a corrector scheme for particle integration in multiphase flow that ensures phase-consistent trajectories.

\section{Related Work}
\label{sec:related-work}

The field of research most related to our work is feature tracking.
Post et al.~\cite{CGF:CGF723} provide a survey on the topic.
Graph representation is a commonly employed approach to convey temporal evolution of features, e.g., in the analysis of combustion simulations~\cite{embeddedSurfaces} or combustion experiments~\cite{Robbins2000Visualization}, and the features can be abstracted by glyphs~\cite{reinders2001}.
Often, features are defined by a threshold.
Bremer et al.~\cite{Bremer2010Analyzing} propose a hierarchy-based approach that alleviates the dependency on predefined thresholds.
For the inspection of features at a given time, a linked-view approach is employed in these methods.
Gu and Wang~\cite{Gu2011Transgraph} compute time-dependent state transition probabilities for volumetric data and visualize a 3D view of the volume together with a 2D graph representation of the transitions.
Grottel et al.~\cite{Grottel:2007:VVA:1313046.1313132} visualize the evolution of molecular clusters on a timeline as an addition to a 3D representation of the molecules to monitor the quality of the clustering.
Laney et al.~\cite{Laney:2006:UST:1187627.1187825} track bubbles and visualize their split and merge behavior as a graph.
Further research on feature tracking concentrates on clustering methods, including the work by Ozer et al.~\cite{10.1109/TVCG.2013.117}, where user-defined feature characteristics are used to determine feature groups.
A survey on graph representations is provided by Wang~\cite{eurovisstar.20151111}.
Our method differs from these approaches, in that we focus on the spatio-temporal aspect of feature representation, where the topology is directly conveyed in the feature volume.

For the analysis of the combustion process in engine simulation, Garth et al.~\cite{Garth2007} proposed a set of visualization methods
that operate on time-varying unstructured grids.
A recent work by Sauer et al.~\cite{Sauer2014JointParticle} utilizes particle and volume data from the simulation runs to track features over longer time intervals.
In our method, we insert particles after the simulation run.
On the one hand, this gives a better control over the particle density and therefore the detail level, on the other hand, however, it necessitates some corrector schemes for phase-consistent advection in multiphase flow, as presented in this paper.

Delocalized quantities are employed in unsteady flows to statically investigate the dynamics of scalar fields~\cite{10.1109/MCG.2008.106}, where quantities are averaged over time along integral curves.
Our concept can be viewed as a special case of this, where only the values at trajectory end points are stored at the seed point.
The visualization of delocalized quantities can be improved using uncertainty information from the FTLE field~\cite{Sadlo2015}.
For the topology analysis in general unsteady flow, the FTLE has become a standard visualization method~\cite{haller2001distinguished}.
Sanderson~\cite{Sanderson2014} proposed an alternative inspired by Lyapunov exponents that is based on traveled particle distance instead of particle divergence.

The dynamics of fluid flow is often represented by surfaces or volumes.
In the work by van~Wijk~\cite{vanWijk1993ImplicitStreamSurfaces}, stream surfaces are 
obtained by extracting isosurfaces from a scalar field. This in turn is generated by advecting streamlines backward up to the boundary with predefined scalar values and resampling those values along the streamlines.
Becker et al.~\cite{Becker:1995:UFV:832271.833882} use flow volumes in unsteady flow to reveal the flow dynamics around regions of interest.
An implicit version of the flow volumes by Xue et al.~\cite{Xue2004RenderingImplicitFlowVolumes} allows for more detailed inspection of the flow.

In our work, the labels corresponding to the volumetric contributions can be interpreted as different materials.
Material interface reconstruction is a challenging topic in visualization, especially with respect to volume preservation in multi-material configurations~\cite{Anderson2010volumeAccurate,Meredith2010analysis,Bonnell03materialinterface}.
For the extraction of volume boundaries, we adhere to the representation by the marching cubes algorithm~\cite{Lorensen:1987:MCH:37402.37422}, since in our case the material distribution within the cells is explicitly given by points with assigned labels.
Moreover, as the number of materials (i.e., labels) can be arbitrarily large for a given investigation, it would complicate the computation of the interface based only on the fraction information.
Further research in material interface analysis includes the work by Obermaier et al.~\cite{Obermaier:2012:VOM}, where the stability of reconstructed interfaces is investigated by comparing it with time surfaces.

For the visualization in multiphase flow, we use a corrector method to ensure that the advected particles remain in the given phase during tracing.
Related to this problem are approaches for surface tracking.
Stam~\cite{Stam2011} proposed a method for the computation of interface velocities to properly translate surfaces.
Bojsen-Hansen et al.\cite{Bojsen-Hansen:2012:TSE:2185520.2185549} developed a method for tracking surfaces undergoing topology changes, without prior information on the underlying physics.
In the work by Bonneel et al.~\cite{Bonneel2011}, Lagrangian transport is used for correct displacement interpolation. Solomon et al.~\cite{Solomon:2015:CWD:2809654.2766963} optimize the transportation in terms of Wasserstein distances, which allows for efficient shape transformation.
In our work, we analyze the evolution of volumes, and computing Wasserstein distances in this case would be computationally prohibitive and also would not guarantee physically correct correspondences.

We exemplify the utility of our technique using multiphase flow simulation datasets.
Investigation of droplets is an active area of research~\cite{yokoi2008,tryggvason2011direct,bothe2012}, and so is the analysis of liquid jets~\cite{Fuster2009550, sussman2010, Ertl2015}.
We refer the reader to Fuster et al.~\cite{Fuster2009} for a detailed introduction to multiphase flow simulation and to Lefebvre~\cite{lefebvre1988atomization} for a thorough description of liquid atomization and sprays.

\section{Simulation Data}
\label{sec:simulation-data}
\begin{figure*}[t]
  \def\clipmethod{30}
  \centering
  \subfigure[]{%
    \includegraphics[height=0.24\linewidth,clip=true,trim=0px \clipmethod px 0px 0px]{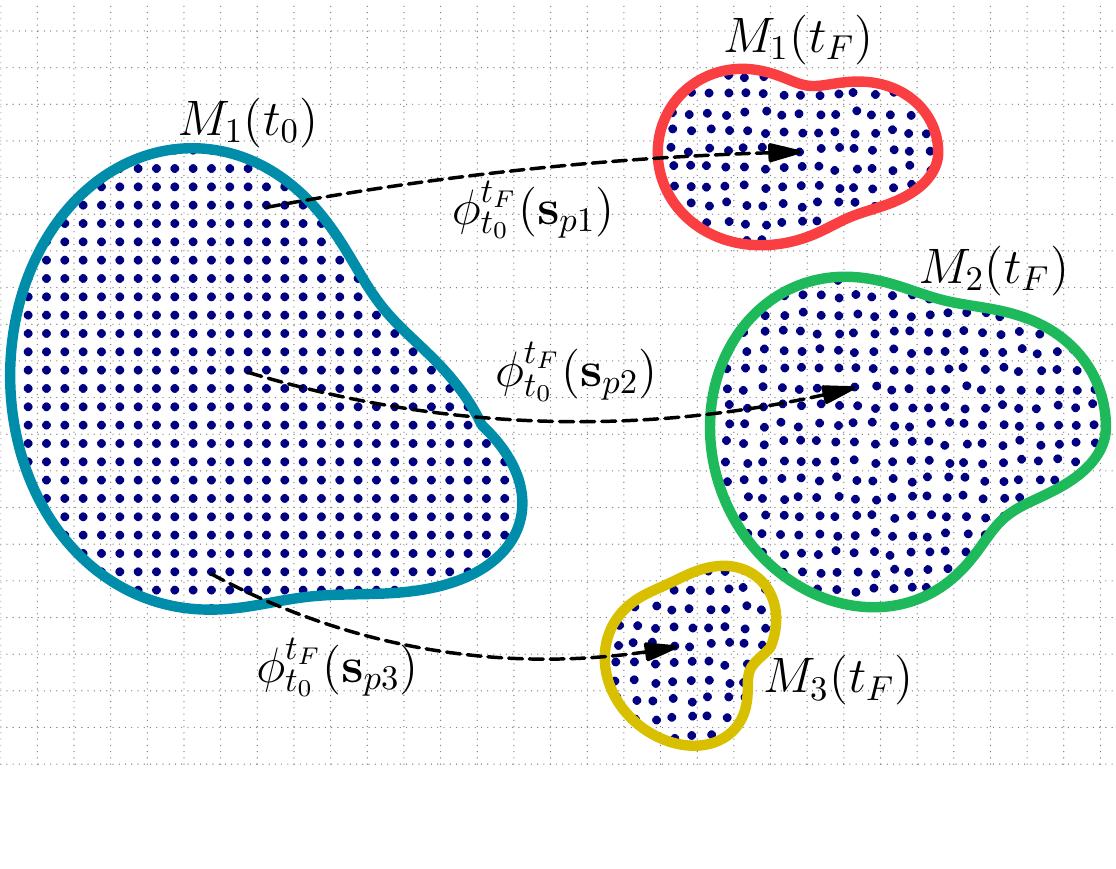}\label{fig:vis-methodstep1}}
  \hfill%
  \subfigure[]{%
\includegraphics[height=0.24\linewidth,clip=true,trim=0px \clipmethod px 0px 0px]{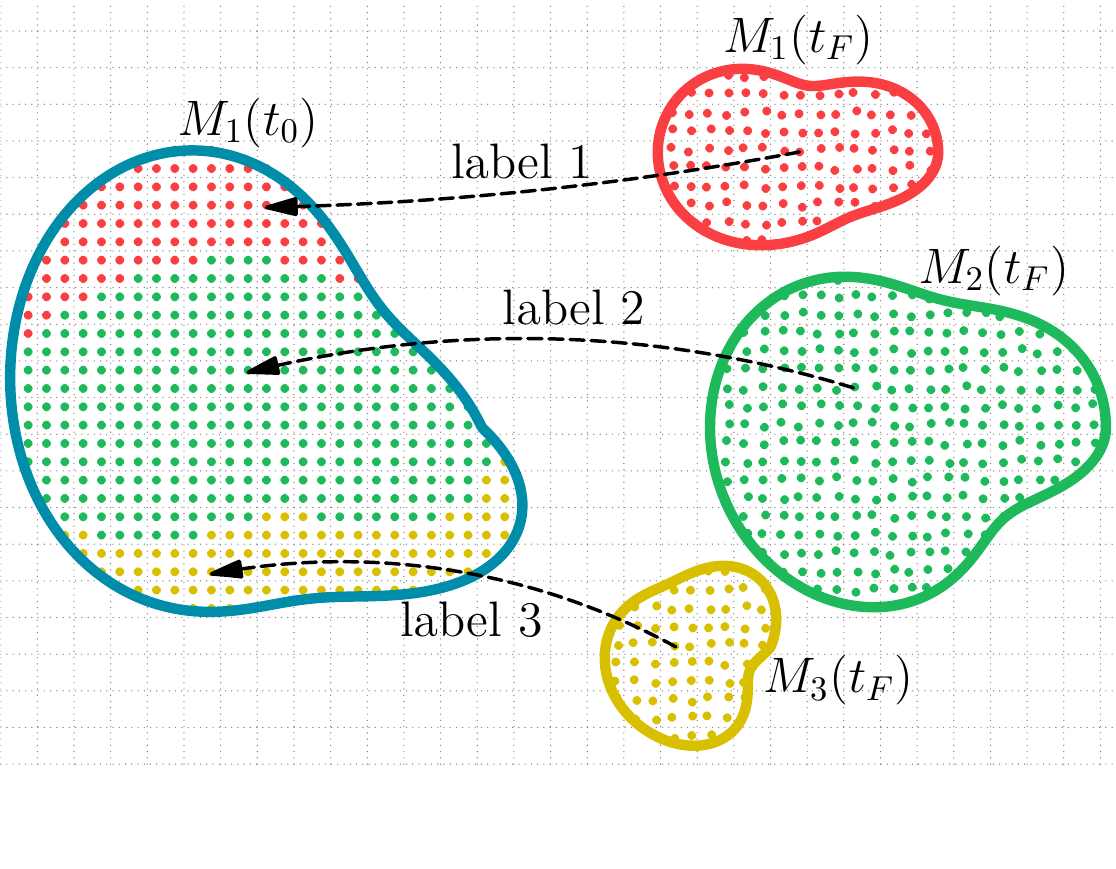}\label{fig:vis-methodstep2}}
  \hfill%
  \subfigure[]{%
\includegraphics[height=0.24\linewidth,clip=true,trim=0px \clipmethod px 0px 0px]{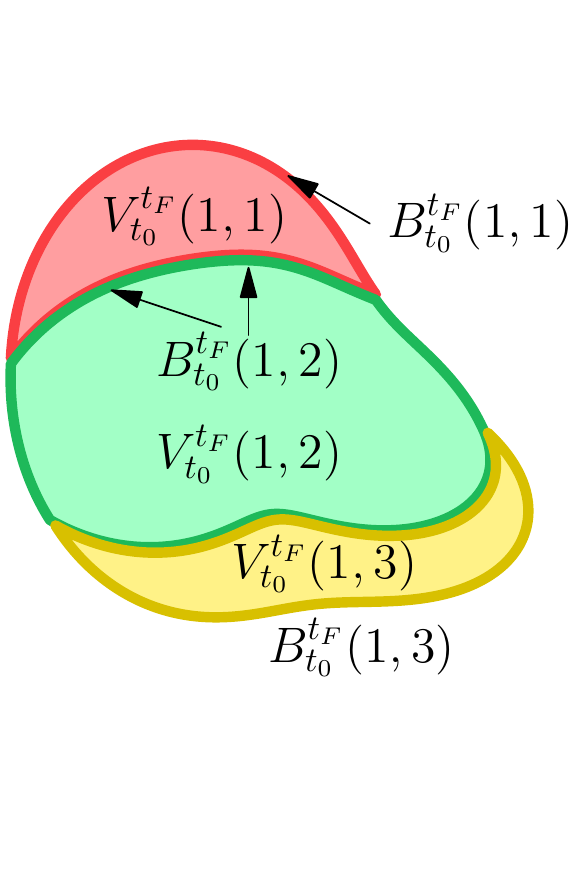}\label{fig:vis-methodstep3}}
  \hfill%
  \caption{
    Visualization of feature separation boundaries.
    There is one feature at time $t_0$, and three features at $t_F = t_2$.
    \subref{fig:vis-methodstep1}~Particles are seeded at time step $t_0$ inside the feature and advected to $t_F$, resulting in a flow map $\phi_{t_0}^{t_F}$ (black arrows).
    At $t_F$, the features have been labeled red, green, and yellow.
    \subref{fig:vis-methodstep2}~Feature labels are assigned to the particles that are inside a given feature.
    These labels are then transferred to the seed points (arrows).
    \subref{fig:vis-methodstep3}~For each label, the corresponding volume $V$ is identified (red, green, and yellow areas), and the boundary $B$ is extracted (corresponding darker curves).
  }\label{fig:vis-methodsteps}
\end{figure*}

The data used in this paper are computational fluid dynamics (CFD) simulations of two-phase (liquid-gas) and two-component (liquid-liquid) flow, based on the Navier-Stokes equations and discretized on rectilinear grids~\cite{Eisenschmidt2016508}.
The velocity is represented by a cell-based vector field $\mathbf{u}(\mathbf{x},t)$.

In case of scientific multiphase flow simulations, the volume of fluid (VOF) method~\cite{Hirt1981201} is typically used for interface tracking, where an additional volume fraction field $f(\mathbf{x},t)$ is maintained for each cell.
A cell contains only the gas phase when $f=0$, only the liquid phase when $f=1$, and the interface is located within cells with $0<f<1$. For liquid-liquid flow, the gas phase is replaced by the surrounding liquid phase.
To maintain sharp interfaces, piecewise linear interface calculation (PLIC)~\cite{young1984plic3D} is used which approximates the interface by planar patches.
In this paper, we use linear interpolation in time for tracking, whereas for sampling of the fraction field, we use the PLIC interface to determine the sampled value. 
Only features that undergo advective transport are considered, since, e.g., diffusive features would be difficult to capture with the particle-based approach, and are therefore beyond the scope of our work.

Our technique employs particle advection in order to determine the volumetric correlation of features between a reference time step $t_0$ and the target time step $t_F$.
Usually, only a certain fraction of timesteps of a simulation are stored for later analysis.
Using these data with reduced time resolution can lead to errors in the estimation of trajectories, especially in multiphase flow, where particles potentially leave the initially assigned phase for this reason.
This problem could be avoided if particles were advected during the simulation, which is increasingly popular, as reported by Sauer et al.~\cite{Sauer2014JointParticle}.
Additionally, particles could be densely populated using the method proposed by Agranovsky et al.~\cite{7013206} in order to capture more details of the topology of feature dynamics.
In our datasets, however, particle data was not provided.
Nevertheless, to still ensure robust particle advection in terms of phase consistency in multiphase flows, we introduce a feature-based corrector method, as described in Section~\ref{sec:phase-consistent-traj}, which utilizes the $f$-field in a corrector step during particle advection to ensure that particles stay in the respective phase throughout integration.

\section{Visualization Method}
\label{sec:vis-method}
We define a feature $M$ as a region with the value of the quantity $f$ exceeding a corresponding threshold $\tau$: $M=\{\vecx : f(\vecx,t) > \tau\}$.
In our experiments, $f$ represents the fraction of fluid in a cell in multiphase flow and $\tau$ is set to $0$, so that the liquid phase (or one of the phases in liquid-liquid flow) represents the features. For other scalar quantities, $\tau$ could be set to a specific value of interest.
Feature separation is illustrated in Figure~\ref{fig:vis-methodstep1}, where an initial feature $M_1$ at time $t_0$ splits within the time interval $]t_0,t_F[$, resulting in three features $M_{\{1,2,3\}}$ at time $t_F$.

\subsection{Separation-Based Feature Segmentation}
\label{sec:boundaries}
With our visualization technique, we want to capture how features develop topologically in the course of time.
For instance, if a feature splits into two new ones, we want to learn how the volume of the initial feature is divided among them, or, in other words, what the volumetric contributions of the new features in the initial one are.
To accomplish this goal, we have to determine the spatio-temporal correspondence between the feature $M_i(t_0)$ at time $t_0$, and features $M_j(t_F)$ at some other time $t_F$.
That is, for each $M_j(t_F)$, we define its volumetric contribution within $M_i(t_0)$ as
\begin{equation}
\label{eq:topo-volume}
V_{t_0}^{t_F}(i,j) = \{\vecx: \vecx \in M_i(t_0) \land \phi_{t_0}^{t_F}(\vecx) \in M_j(t_F)\},
\end{equation}
where $\phi_{t_0}^{t_F}(\vecx)$ maps the initial position $\vecx$ 
at time $t_0$ to its position at time $t_F$, as it is advected by the flow.
We refer to $\Dt$ = $t_F - t_0$ as the computation time interval.
For our visualization, we extract the closed boundary of the volume $V$:
\begin{equation}
\label{eq:topo-boundary}
B_{t_0}^{t_F}(i,j) = \partial V_{t_0}^{t_F}(i,j).
\end{equation}
Note that for $\Dt=0$, $B$ bounds the whole feature volume if $i=j$, and $B$ does not exist if $i\neq j$.
Also, note that our method allows $t_F$ to be earlier in time than $t_0$ if we set $\Dt < 0$.
In Figure~\ref{fig:vis-methodsteps}, the method is illustrated for a simple case where a feature splits into three.

Volumetric contributions $V_{t_0}^{t_F}(i,j)$ can be composed of disconnected segments, either due to a merge followed by a split of the initial features, or due to disjoined volume segments inside the initial feature that together form a separate feature.
As will be shown in Section~\ref{sec:boundary-extraction}, our visualization method allows one to readily identify and analyze these cases.

\subsubsection*{Separation Surfaces}
\label{sec:s-surfaces}
\begin{figure}[t]
  \centering
  \subfigure[]{\includegraphics[height=0.38\linewidth]{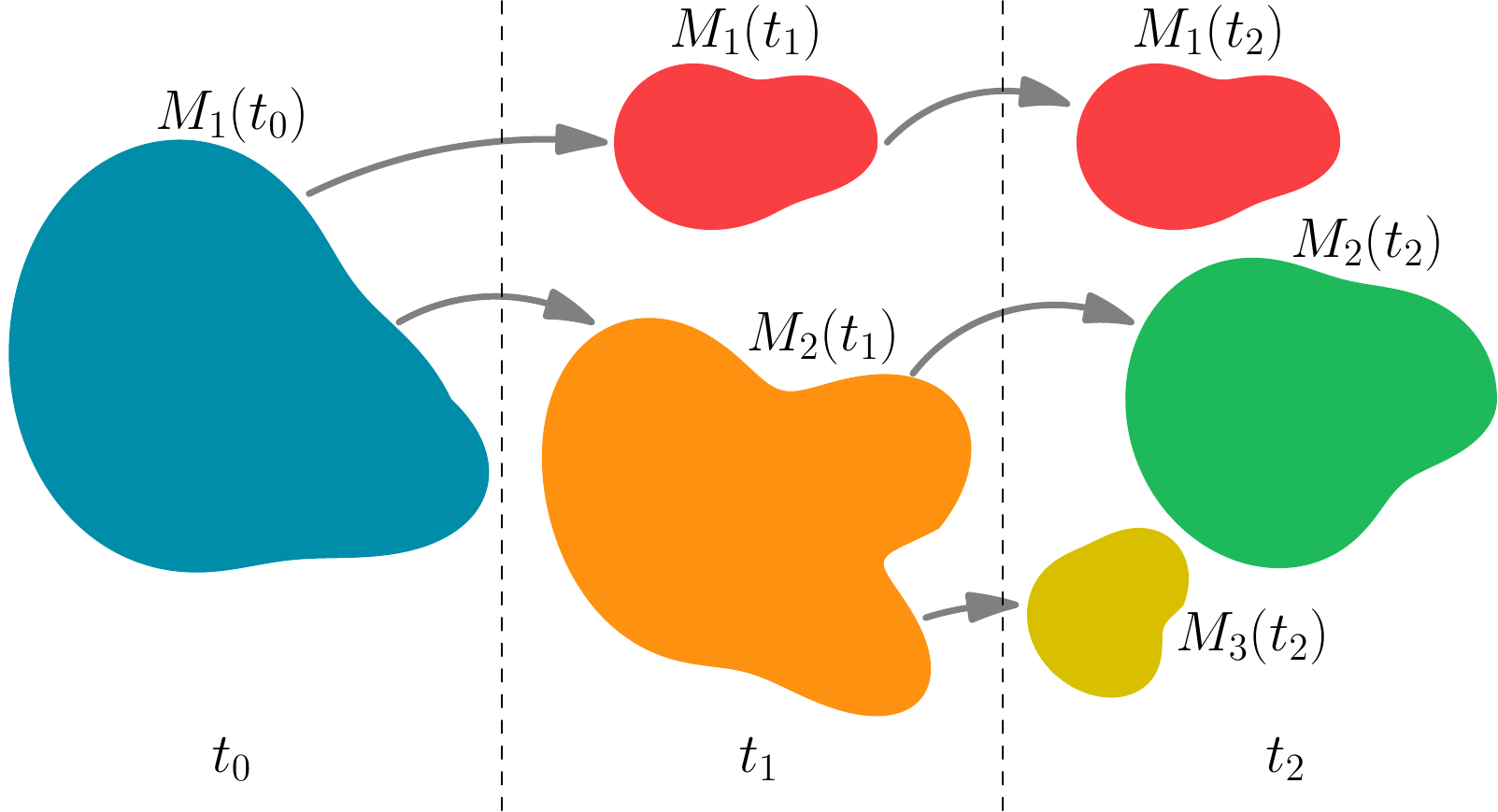}
    \label{fig:vis-ss-methodstep2}}
  \hfill%
  \subfigure[]{\includegraphics[height=0.38\linewidth]{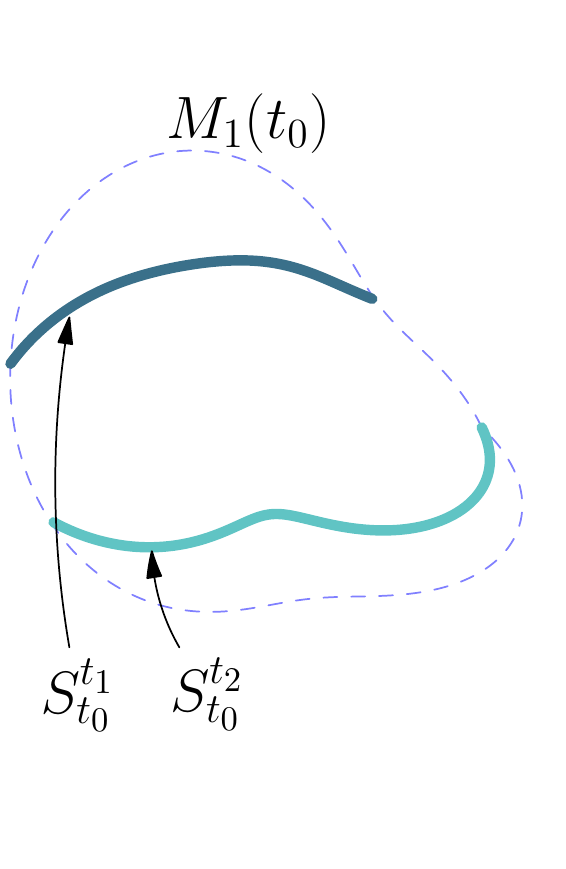}
    \label{fig:vis-ss-methodstep3}}
  \caption{
    Visualization of separation surfaces.
    \subref{fig:vis-ss-methodstep2}~There is one feature at time $t_0$, which splits into the red and the orange feature at the intermediate timestep $t_1$, and the orange one splits again at $t_2 = t_F$ into green and yellow feature. 
    \subref{fig:vis-ss-methodstep3}~Temporal separations surfaces $S_{t_0}^{t_1}$ and $S_{t_0}^{t_2}$ partition the initial feature according to the volumetric correspondences of the newly created features at time $t_1$ and $t_2$, respectively.
  }\label{fig:vis-ss-methodsteps}
\end{figure}
To support temporal analysis of feature separation, we supplement our technique with the extraction of temporal separation surfaces, denoted $S$. 
While the above method finds the volumetric correspondences in the initial time $t_0$ to the features at target time $t_F$, here, the separation surfaces are constructed within the whole interval $]t_0,t_F]$. They divide the volumetric contributions in the initial feature as the corresponding features split into new features.

This is illustrated in Figure~\ref{fig:vis-ss-methodstep2}, where the initial feature $M_1(t_0)$ has split into two at time $t_1$, and the resulting feature $M_2(t_1)$ has further separated into new features $M_2(t_2)$ and $M_3(t_2)$ at time $t_2$. 
In Figure~\ref{fig:vis-ss-methodstep3}, both split events are illustrated by the separation surfaces $S_{t_0}^{t_1}$ and $S_{t_0}^{t_2}$ that divide the volume segment according to the volumetric contributions of the feature resulting from the separation.

To create the separation surfaces and determine the time at which they occur, changes in contributions are detected within small time intervals.
Specifically, for each time increment $\delta t = t_2 - t_1$, where $\delta t \ll  \Dt$, contributions $V_{t_0}^{t_2}(i,k)$ in each feature $M_i(t_0)$ are computed and compared with the previous contributions $V_{t_0}^{t_1}(i,j)$. 
If the number of contributions $V_{t_0}^{t_2}(i,k)$ for which $V_{t_0}^{t_1}(i,j) \cap V_{t_0}^{t_2}(i,k) \neq \emptyset$ is greater than one, a separation has occurred in the given interval inside the volume contribution $V_{t_0}^{t_1}(i,j)$, and the separation surface is generated in $M_i(t_0)$ where different $V_{t_0}^{t_2}(i,k)$ adjoin.

Theoretically, the temporal evolution of feature separation could be accomplished by repeatedly extracting the closed boundaries $B_{t_0}^{t_k}(i,j)$ for varying $k$.
This would, however, lead to repeated construction of overlapping boundaries that would be difficult to analyze.
The separation surfaces $S$, on the other hand, provide an open structure that complements the extracted boundaries $B$.

\begin{figure}[t]
  \centering
  \includegraphics[width=0.8\linewidth]{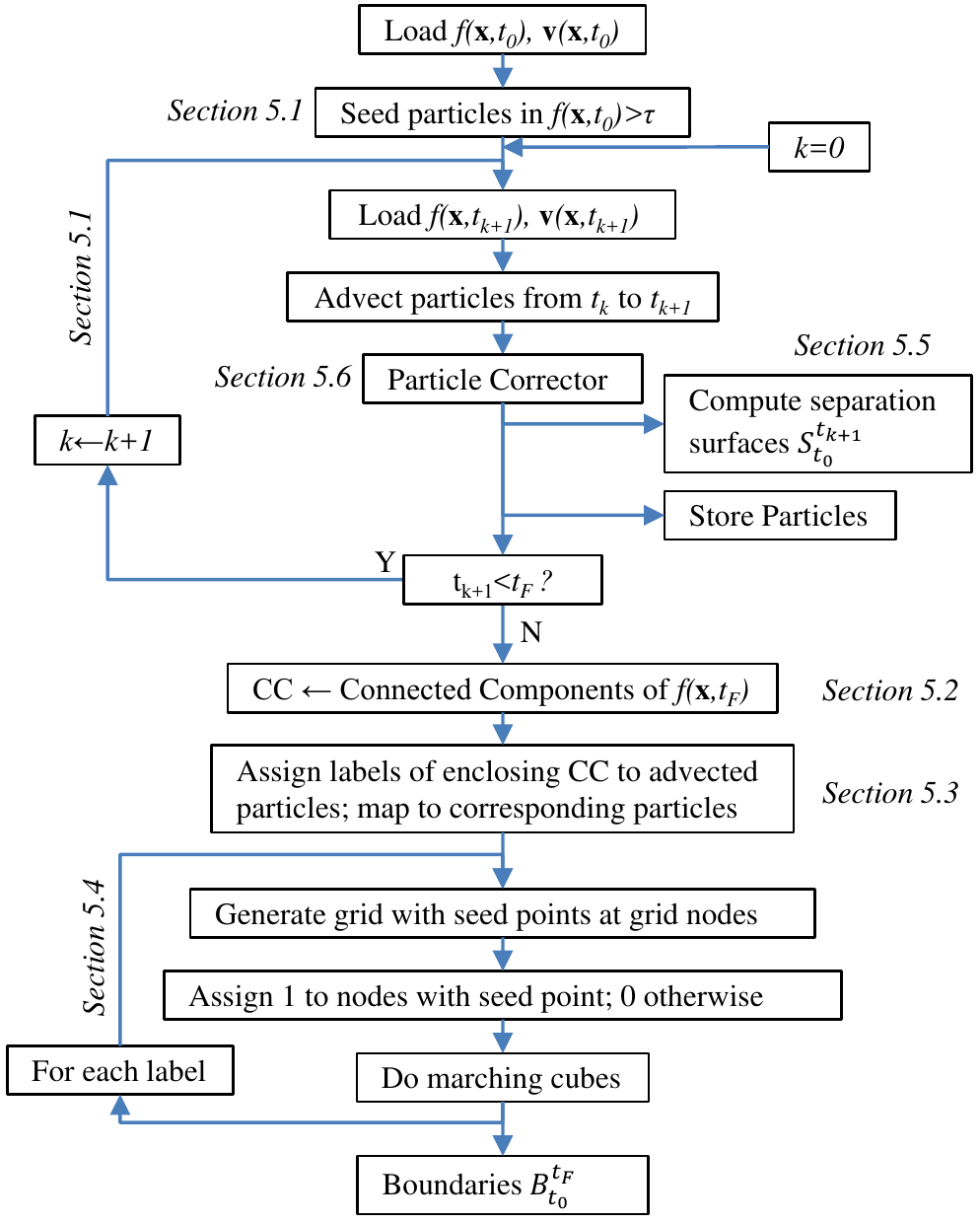}
  \caption{
Overview diagram of the numerical approach, with section numbers given for relevant stages.
The upper loop represents advection performed between consecutive simulation time steps and includes storage of advected particles used for visualization. Separation surfaces $S$ reveal temporal changes in feature segmentation and therefore are computed between consecutive simulation steps.
The lower loop is performed to obtain boundary $B$ for each label, i.e., for each feature at target time $t_F$.
  }\label{fig:method-overview}
\end{figure}

\section{Numerical Approach}
\label{sec:numerical-approach}
\begin{figure*}[t]
  \centering
  \hfill%
  \subfigure[]{\includegraphics[width=0.2\linewidth]{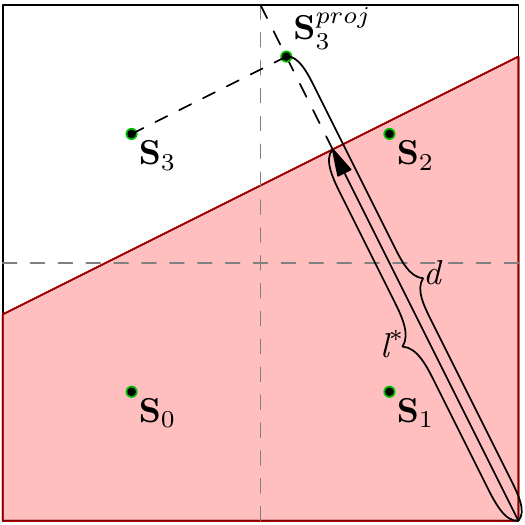}
    \label{fig:plic-seeding}}
  \hfill%
  \subfigure[]{\includegraphics[width=0.2\linewidth]{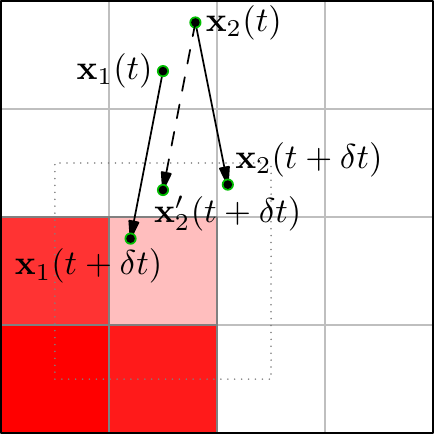}
    \label{fig:smart-corrector}}
  \hfill%
  \subfigure[]{\includegraphics[width=0.2\linewidth]{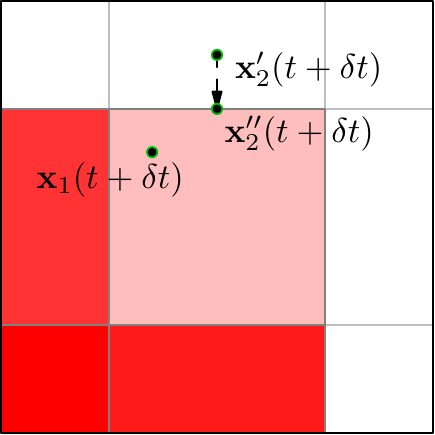}
    \label{fig:vof-corrector}}
  \hfill%
  \subfigure[]{\includegraphics[width=0.2\linewidth]{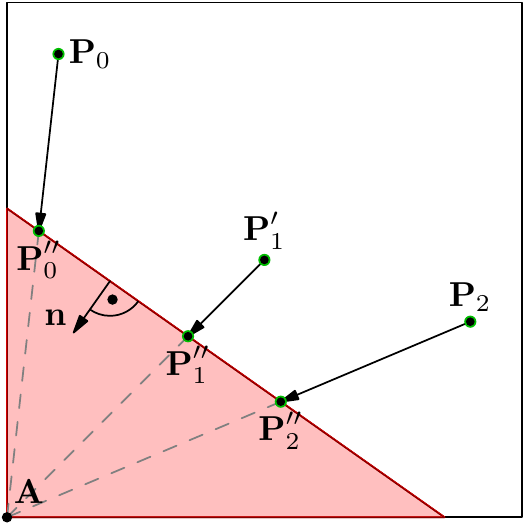}
    \label{fig:plic-corrector}}
  \hfill%
  \caption{
    Particle seeding and corrector method in multiphase flows with PLIC interface reconstruction.
    \subref{fig:plic-seeding}~Seed positions $\vec{s}_\particle$ are determined in each cell with $f>0$ by the refinement parameter $r$.
    Here $r=1$, and $\vec{s}_{1}$--$\vec{s}_{4}$ are at the centers of four subcells.
For each $\vec{s}$, the $\vec{n}$-projected distance $d$ to the attachment point $\vec{a}$ (i.e., the last point swept by the PLIC-plane moving along the normal) is compared with the PLIC patch translation $l$ to determine if particle $p$ is seeded at $\vec{s}_p$.
$\vec{s}_{4}$ will not be seeded in this case.
\subref{fig:smart-corrector}--\subref{fig:plic-corrector}~Three stages of particle correction.
    \subref{fig:smart-corrector}~After each advection step, it is checked whether some particles left the original phase ($f>0$), and if so, the position is computed with the displacement vector of the nearest phase-consistent particle. \subref{fig:vof-corrector}~(Dotted rectangle from~\subref{fig:smart-corrector}) If still not in cell with $f>0$, the particle is moved in the direction of the nearest such cell up to its boundary.
\subref{fig:plic-corrector}~Afterwards, the particles outside the volume enclosed by PLIC patch are projected onto the PLIC plane in the direction of $\vec{a}$.
  }\label{fig:corrector}
\end{figure*}
In our method, we operate on both the Eulerian frame, in which the simulation data is defined, and the Lagrangian frame, where the inserted particles represent the feature volume and allow us to determine the volumetric correspondences.

Our visualization framework works as follows.
The input to our approach are time series of a vector field $\vec{u}$ for particle advection and a scalar field for feature definition.
The initial time $t_0$ is chosen, at which particles are seeded within $M_i(t_0)$.
These particles are then advected up to time $t_F$, whereby integration is performed between consecutive simulation steps, and so is the construction of separation surfaces.
At time $t_F$, the connected components that define the features $M_j(t_F)$ are determined, and for each advected particle, the surrounding feature is found.
The feature labels are then mapped to the seed points, and finally, boundaries are extracted for each label within $M_i(t_0)$. 
In Figure~\ref{fig:method-overview}, the algorithm is presented as a diagram.
In the following sections, each step, marked on the diagram with the corresponding section number, is described. Please see also the supplementary material for a detailed pseudo-code of the algorithm.

\subsection{Particle Seeding and Advection}
At time $t_0$, particles are seeded at $\vec{s}_{\particle}$ if $f(\vec{s}_{\particle},t_0) > \tau$.
This is illustrated in Figure~\ref{fig:plic-seeding}, where $\tau$ is represented by the PLIC interface that divides the cell into liquid and gaseous phase, and seed points are located only at positions $\vec{s}_\particle$ enclosed by the PLIC patch. 

To determine if $\vec{s}_\particle$ lies within the phase of interest, we compute the seed position relative to the PLIC patch.
Therefore, we have to compute the patch orientation and position.
We first compute the gradient of the $f$-field at the cell center $\vecx_c$, and then normalize
it to obtain the PLIC normal $\vec{n}$.
Next, we determine the attachment point $\vec{a}$ as the most distant cell node from the PLIC patch in the direction of $\vec{n}$.
The PLIC patch position $l$, i.e., the distance from $\vec{a}$, is obtained iteratively by ensuring that the volume enclosed by the patch equals the value $f$ in the given cell.
Then, we compare $l$ with the distance $d$ from $\vec{s}$ projected onto the normal.
If $d < l$, a particle is positioned at $\vec{s}_{\particle}$.
In the figure, the distance $d>l$ for $\vec{s}_{m+3}$ implies wrong phase, and hence, no particle will be seeded there.
Please note that for anticipated application to different data models, trilinear interpolation can be used instead of PLIC reconstruction to determine the threshold position.

The seed position $\vec{s}_{\particle}$ and the maximum number of seeds per cell $n_c$ are controlled by a global refinement parameter $r$ in the form \smash{$n_c = (2^d)^r$}, where $d$ is the data dimension ($d=3$ in our datasets).
For $r=0$, the seeds are positioned at the centers of the simulation cells.
For $r>0$, the cells are recursively divided $r$ times into equally sized subcells, and the particles are positioned at the centers of these subcells. 

To compute the flow map \smash{$\phi_{t_0}^{t_F}(\vec{s}_{\particle})$}, we set the particles $\particle$ at the seed points and advect them up to time $t_F$ (Figure~\ref{fig:vis-methodstep1}) using the standard fourth-order Runge-Kutta scheme performed between consecutive simulation time steps in the interval $[t_0,t_F]$.
The seed points $\vec{s}_{\particle}$ are stored for later processing to determine the correspondence between features at times $t_0$ and $t_F$.

\subsection{Feature Labeling}
\label{sec:feature-labeling}
To extract the features at time $t_F$, we first create a grid with the same structure as the input simulation grid.
We equip this grid with a bit mask, where cells $c$ with $f(\vecx_c,t)>\tau$ are set to $1$ and the others to $0$.
With this bit mask, we can find connected components that correspond to the features $M_j(t_F)$.
To this end, we employ region growing.
Here, two feature cells $c_m$ and $c_n$ are considered connected if they are face neighbors.
Each feature is then identified by a label $j \in \mathbb{N}$.
The connected components are stored in a label field, which again structurally corresponds to the simulation grid.
This greatly simplifies particle assignment, as described in the next section.
All grid cells without connected components are marked with an invalid label (i.e., $-1$).
In Figure~\ref{fig:vis-methodstep1}, the labeled features are marked red, green, and yellow.
Feature labeling can be performed independently of the advection step, since determination of feature labels only requires the scalar field $f(\vecx,t_F)$.

\subsection{Particle Label Assignment}
\label{sec:particle-labels}
In the next step, we have to determine those particles $\particle$ for which \smash{$\phi_{t_0}^{t_F}(\vec{s}_{\particle}) \in M_j(t_F)$}.
For each advected particle $\particle$ at time $t_F$, we determine by which feature $M_j$ it is bounded (Figure~\ref{fig:vis-methodstep2}) and assign the corresponding label $j$ to the particle.
If it does not lie inside any feature, a non-valid label (i.e., $-1$) is assigned to it.
Since we operate on rectilinear grids, we can find the bounding cell by iterating over the node coordinates $x_i$ of the grid until $x_i < x_{\particle} < x_{i+1}$, with $x_i\in\{x,y,z\}$.
The feature label $j$ is then mapped to the stored seed points $\vec{s}_{\particle}$ corresponding to the advected particles (shown by dashed arrows in Figure~\ref{fig:vis-methodstep2}).
Additionally, to visualize the temporal evolution of the features, the intermediate particle positions are stored during the computation of the visualization, and the computed labels are assigned to them at this stage.
For scalar quantities beyond multiphase flow, the value $f$ at the particle position can be trilinearly interpolated.
If the particle is inside a cell without a label but the trilinear interpolation results in $f > \tau$, we can compute the gradient of $f$ and assign the label of the nearest cell with $f > \tau$ in gradient direction.

\subsection{Boundary Extraction}
\label{sec:boundary-extraction}
The labeled particles allow us to extract the volume boundaries \smash{$B_{t_0}^{t_F}$} within the feature \smash{$M_i(t_0)$} that correspond to the features \smash{$M_j(t_F)$}, according to Equation~\ref{eq:topo-boundary}.
For each label $j$, we find all seed points $\vec{s}_{\particle}$ with this label and compute a bounding box around them.
Subsequently, a rectilinear grid is generated within the bounding box such that the seed points coincide with the grid nodes.
At each grid node, we set a value $1$ if a seed point is actually located at this position, and $0$ otherwise.
This way, we can employ the standard marching cubes algorithm to extract the boundaries.
Since we require that the grid node positions correspond with the seed points, the size and number of cells is implicitly controlled by the refinement parameter $r$.
In Figure~\ref{fig:vis-methodstep3}, the boundaries $B$ are marked by darker curves that bound the volumes $V$.

The disconnected feature segments (Section~\ref{sec:vis-method}) can be detected either directly from the color of the boundaries (each feature maps to a different color label of the volume regions, and disconnected volume regions belonging to the same feature have the same color) or using the particles from intermediate time steps that also carry the feature label information.

\subsection{Extraction of Separation Surfaces}

To extract the separation surfaces $S$ described in Section~\ref{sec:s-surfaces}, after each advection substep, feature labeling (Section~\ref{sec:feature-labeling}) and particle label assignment (Section~\ref{sec:particle-labels}) are performed to obtain the current feature segmentation.
The resulting seed labels are stored for two subsequent time steps $t_k$ 
and $t_{k+1}$.
For each label at $t_k$, the corresponding seeds and their labels at $t_{k+1}$ are analyzed. 
More than one unique label at $t_{k+1}$ indicates that the segment has split in the current time interval, and therefore, separation surfaces are extracted by performing a modified marching cubes algorithm for each $2$-combination of the labels.
In the algorithm, node values are set to ``$+$'' and ``$-$'' which correspond to either of the labels, and the surface passes through the edge centers between ``$+$'' and ``$-$'' nodes. 
To ensure open surfaces, the outer nodes (i.e., not belonging to any of the two labels) are marked with an invalid negative value.
In the final step of marching cubes, the edges of each triangle are tested for whether they lie on edge with the outer node, and discarded in this case.
Finally, the surfaces and the corresponding time stamps $t_{k+1}$ are stored for visualization.

\subsection{Phase-Consistent Trajectories in Multiphase Flow}
\label{sec:phase-consistent-traj}

Since typically only a fraction of simulation time steps is saved for analysis and visualization (due to potentially high storage demands of large simulations), and because multiphase flow is highly nonlinear (due to, among others, surface tension forces), there is usually not enough information to ensure phase consistency of the advected particles, i.e., the particles may stray off to the other phase during advection.
However, assuming that physically-based phase transitions do not occur in the simulation (which is the case in our applications), the particles must stay in the same phase throughout advection.
To accommodate this requirement, and to provide plausible results with the available information, we introduce a three-stage approach that utilizes the scalar field representing the phase of interest to correct positions of stray particles during advection.
\subsubsection*{Particle Corrector}
\label{sec:vof-corrector}

After each integration step between $t_k$ and $t_{k+1}$, we test each advected particle if it remained in the assigned phase.
If this is not the case, a three-stage procedure is applied to translate the stray particle back to its original phase.
In the first stage, the neighborhood of the particle is searched for valid particles (i.e., particles that remain in the assigned phase at $t_{k+1}$).
That is, in the $3\times 3 \times 3$ cell neighborhood at time $t_k$ (i.e., before the current advection step) the nearest neighbor (if any) is chosen. Its displacement vector $\mathrm{d}\vec{x}_p=\vec{x}_p(t_{k+1})-\vec{x}_p(t_k)$ is applied to the stray particle, i.e., $\vec{x}'(t_{k+1}) = \vec{x}(t_k)+\mathrm{d}\vec{x}_p$ (Figure~\ref{fig:smart-corrector}).
Since this does not guarantee that the particle will be back in the original phase, 
in the second stage, the nearest cell with $f>0$ is searched (using a loop over neighboring cells), and the particle is translated to the boundary of this cell along the line connecting the particle with the cell center, as shown in Figure~\ref{fig:vof-corrector}, where the new particle positions are denoted $\vecp''$.
Afterwards, in the third stage, for each particle in the interface cell, the position of the PLIC patch is calculated at the current simulation time step.
Then, it is tested if the particle position lies within the volume enclosed by the PLIC patch.
If not, the particle is translated along the line $\vecp-\vec{a}$ to the patch (Figure~\ref{fig:plic-corrector}).

\section{Implementation}
\label{sec:implementation}

\begin{figure*}
  \centering
  \begin{minipage}[b]{.15\linewidth}   
    \includegraphics[width=0.98\linewidth]{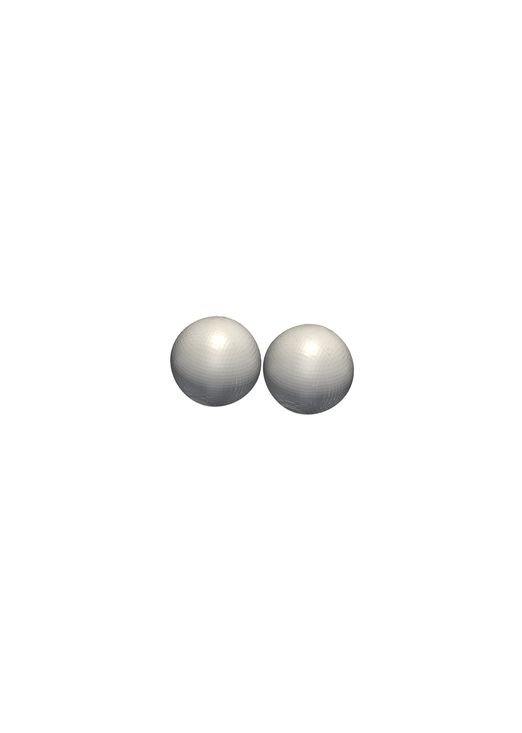}
    \\
    \subfigure[$0.478\,\textrm{ms}$]{%
      \includegraphics[width=0.98\linewidth]{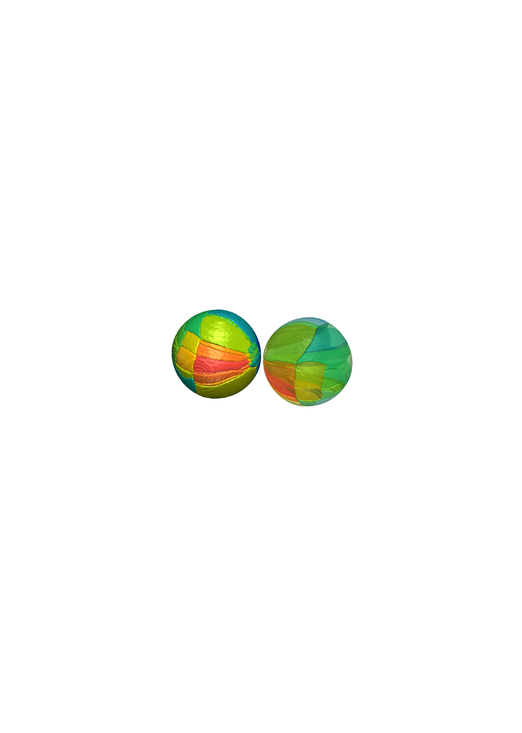}
    \label{fig:teaser07}}
  \end{minipage}
  \hfill%
  \begin{minipage}[b]{.15\linewidth}   
    \includegraphics[width=0.98\linewidth]{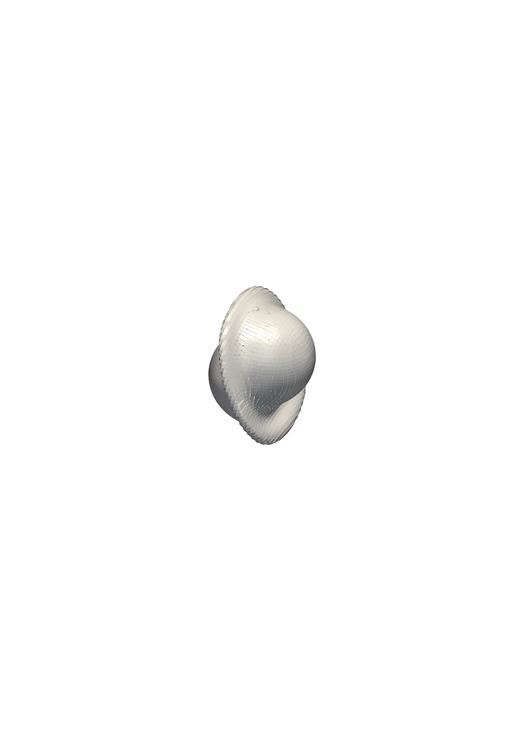}
    \\
    \subfigure[$0.719\,\textrm{ms}$]{%
      \includegraphics[width=0.98\linewidth]{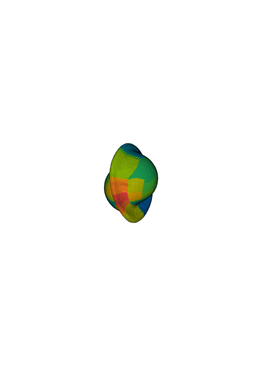}
    \label{fig:teaser08}}
  \end{minipage}
  \hfill%
  \begin{minipage}[b]{.15\linewidth}   
    \includegraphics[width=0.98\linewidth]{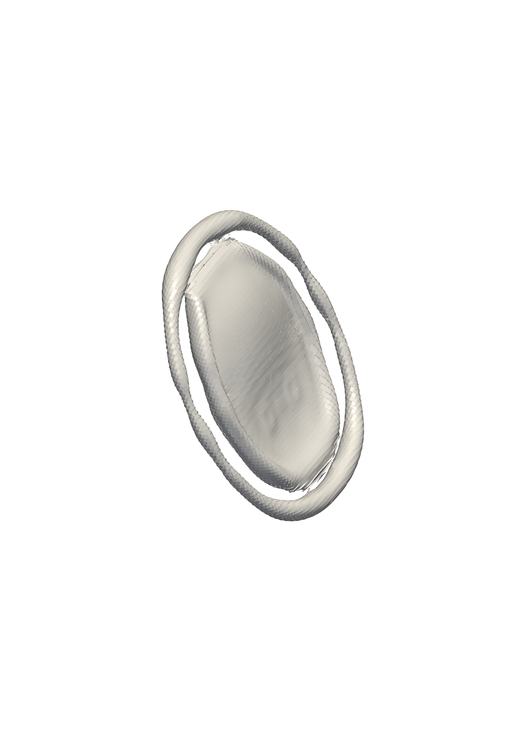}
    \\
    \subfigure[$1.31\,\textrm{ms}$]{%
      \includegraphics[width=0.98\linewidth]{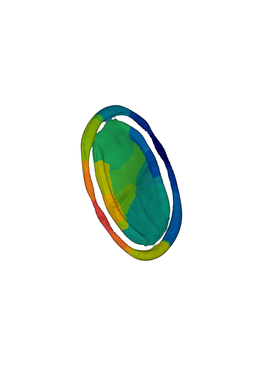}
    \label{fig:teaser09}}
  \end{minipage}
  \hfill%
  \begin{minipage}[b]{.15\linewidth}   
    \includegraphics[width=0.98\linewidth]{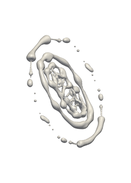}
    \\
    \subfigure[$1.93\,\textrm{ms}$]{%
      \includegraphics[width=0.98\linewidth]{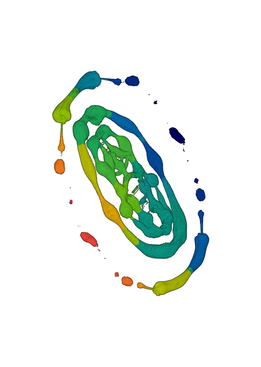}
    \label{fig:teaser10}}
  \end{minipage}
  \hfill%
  \begin{minipage}[b]{.15\linewidth}   
    \includegraphics[width=0.98\linewidth]{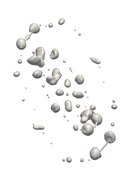}
    \\
    \subfigure[$2.43\,\textrm{ms}$]{%
      \includegraphics[width=0.98\linewidth]{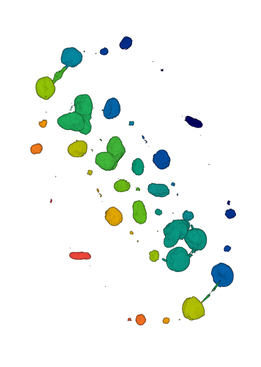}
    \label{fig:teaser11}}
  \end{minipage}
  \hfill%
  \begin{minipage}[b]{.15\linewidth}   
    \includegraphics[width=0.98\linewidth]{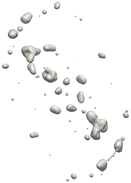}
    \\
    \subfigure[$2.85\,\textrm{ms}$]{%
      \includegraphics[width=0.98\linewidth]{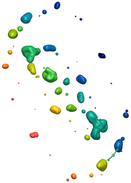}
    \label{fig:teaser12}}
  \end{minipage}
  \hfill%
  \caption{
    Visualization of feature separation in case of two-phase flow with two peripherally colliding drops.
    The top row shows the extracted liquid interface for selected simulation time steps.
    After collision, the drops initially form a disk that disintegrates into many differently shaped droplets.
    In the bottom row on the left, volumetric contributions of each droplet from the last shown simulation time step are visualized, colored according to the labels of these droplets (shown on the right).
    The visualization reveals the topology of the drop disintegration spatially.
    For the four time steps in the middle, particles with colors corresponding to the drop labels are visualized, which allows for spatio-temporal tracking of the features.
  }\label{fig:teaser}
\end{figure*}

We have implemented our visualization method as a plugin in the ParaView visualization framework~\cite{ParaViewGuide}.
To improve the visual representation of the boundaries $B$ and separation surfaces $S$, we smooth the meshes extracted by the marching cubes algorithm.
Although some error is introduced in the process, a smoothed representation improves perception, and hence we consider it a reasonable trade-off.
For the visualization of the labeled features in multiphase flow datasets, we extract the geometric representation of the PLIC interfaces using the method by Karch et al.~\cite{karch2013PLIC}
and resample the labels on the PLIC patches.
Apart from the boundaries we also store intermediate particle positions after every $n$th simulation time step ($n=8$ in our applications), in order to visualize temporal evolution of the features, see for instance Figure~\ref{fig:teaser}.

For the second stage of the particle corrector scheme (translation to a cell boundary), we adopted the line-box intersection algorithm proposed by Kay and Kajiya~\cite{Kay:1986:RTC:15886.15916}.

\subsection{Parallelization}
\label{sec:parallelization}
\begin{figure}[t]
  \centering
  \subfigure[]{\includegraphics[width=0.89\linewidth]{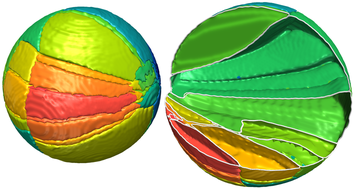}
    \label{fig:dropcoll-cross-sec}}
  \subfigure[]{\includegraphics[width=0.89\linewidth]{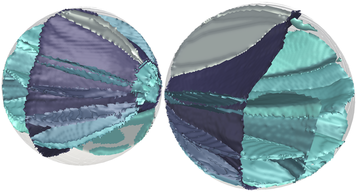}
    \label{fig:dropcoll-s-surfaces}}
  \subfigure[]{\includegraphics[width=0.6\linewidth]{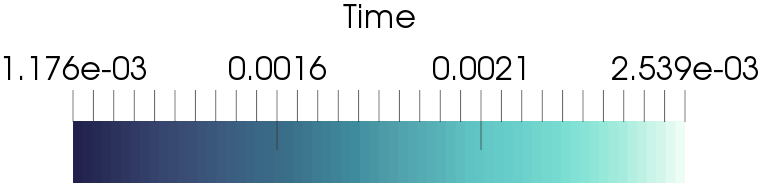}} 
  \caption{
    Visualization of feature separation in \dropcoll/ dataset.
    \subref{fig:dropcoll-cross-sec}~The section cutout (marked with white curve) of the boundaries $B$ from Figure~\ref{fig:teaser} exposes the inner topology structure---almost all developed drops radially expand from the collision center.
    \subref{fig:dropcoll-s-surfaces}~The separation surfaces reveal segmentation in the temporal context. Dark blue surfaces indicates the early separation of disk shaped and ring shaped droplets. Afterwards, both parts separate in similar time further, as indicated by turquoise color. 
  }
\end{figure}

For large datasets, parallelization is necessary due to large memory requirements of the simulation data.
For instance, in the \jet/ simulation (Section~\ref{sec:results-jet}), each simulation time step consists of over $20\,\textrm{GiB}$, which could not be processed on a regular desktop computer.
Hence, we employ data parallelism (i.e., the approach adopted by ParaView), where the data is split among several processes (possibly running on different machines), and each process works on a preassigned subdomain.
In this approach, explicit communication is necessary for data exchange.

Due to the global nature of particle advection, those particles that leave a subdomain must be sent to the processes managing the subdomains the particles have entered.
To avoid cases where the Runge-Kutta substeps are computed across neighboring subdomains, we store ghost cells around each subdomain.
Another global algorithm we employ in our technique is connected component labeling.
Our implementation is based on the method proposed by Harrison et al.~\cite{EGPGV:EGPGV11:131-140}.
Additionally, the transfer of particle labels to their respective seed points must be handled explicitly.
That is, for each particle we store its id within the original subdomain, and the process id responsible for that subdomain.
This information is then used to transfer the labels to the correct processes and to save the label at the correct position in the label array.

In addition to the process-level parallelization, we also employ thread-level parallelism for particle advection, i.e., we employ the OpenMP~\cite{openmp2015} parallel-for loops to speed up iteration over particles being advected.

\section{Results}
\label{sec:results}

We demonstrate the utility of our method on three datasets from direct numerical simulations of incompressible flow.
The \dropcoll/ and \jet/ datasets are two-phase flow simulations, where liquid and gas phases occur simultaneously. The \oil/ dataset contains oil inclusions in water surrounding. For \dropcoll/ and \oil/, $r=2$, whereas for \jet/, $r=0$.

\subsection{Drop Collision}
\label{sec:results-drop-coll}

\begin{figure}[t]
  \centering
  \hfill%
  \begin{minipage}[b]{.45\linewidth}   
    \subfigure[]{\includegraphics[width=0.8\linewidth]{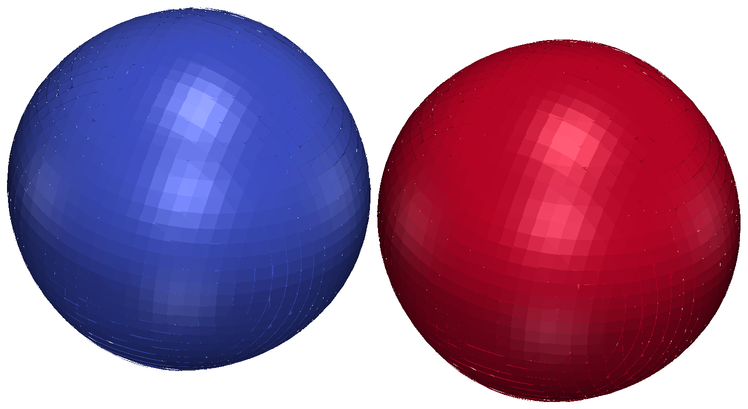}
      \label{fig:dropcoll-reverse1}}
    \subfigure[]{\includegraphics[width=0.8\linewidth]{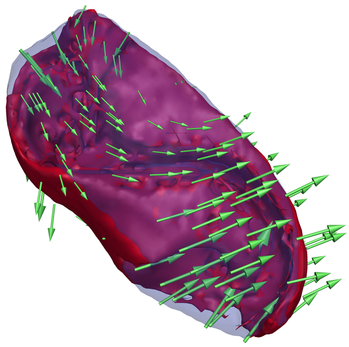}
      \label{fig:dropcoll-reverse3}}
    \hfill%
  \end{minipage}
  \begin{minipage}[b]{.45\linewidth}   
    \subfigure[]{\begin{overpic}[width=0.98\linewidth]{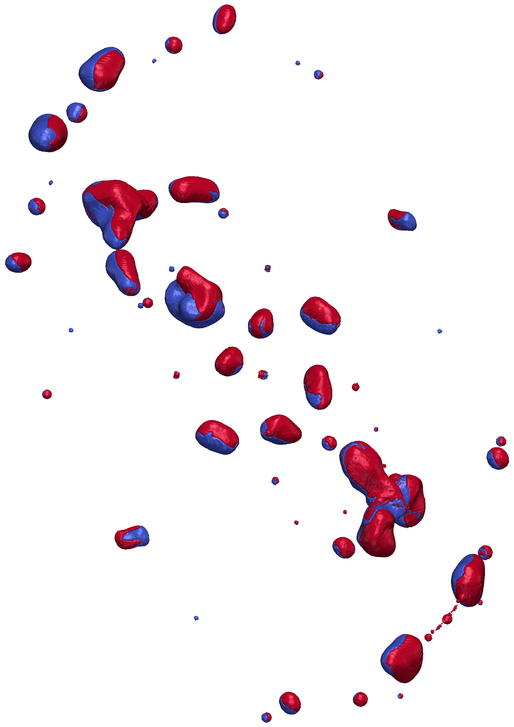}
        \linethickness{0.75pt}
        \multiput(25,35)(3.33,0){3}{\color{black}\line(1,0){1.666}}
        \multiput(25,45)(3.33,0){3}{\color{black}\line(1,0){1.666}}
        \multiput(25,35)(0,3.33){3}{\color{black}\line(0,1){1.666}}
        \multiput(35,35)(0,3.33){3}{\color{black}\line(0,1){1.666}}
      \end{overpic}
      \label{fig:dropcoll-reverse2}}
  \end{minipage}
  \hfill%
  \caption{
    Topology of drop dynamics computed backward in time for the \dropcoll/ dataset.
    \subref{fig:dropcoll-reverse1}~Drops at $t_F$ are labeled (blue and red).
    \subref{fig:dropcoll-reverse2}~The volumetric contributions are visualized at $t_0$.
    \subref{fig:dropcoll-reverse3}~Our technique for selected rotating drop (marked with box in~\subref{fig:dropcoll-reverse2}) reveals S-shaped structure within the drop volume.
  }\label{fig:dropcoll-reverse}
\end{figure}

The first dataset is a simulation of a two-phase flow with two water droplets that collide peripherally.
This dataset consists of $256^3$ cells, which cover a domain size of $1\,\text{cm}^3$, and $461$ time steps.
In the top row of Figure~\ref{fig:teaser}, selected simulation time steps (from $t_0$ to $t_F$) are shown.
We analyze the features from time $t_0=0.478\,\text{ms}$ (i.e., $60$th time step, just before the collision) up to time $t_F=2.85\,\text{ms}$ (i.e., $461$st time step, with small droplets resulting from the collision).
The two merged drops form a flat shape that splits into an inner disk and an outer ring.
Finally, both structures separate into small droplets.

Our visualization in the bottom row reveals the volumetric correspondences between the volumes of the two initial drops and the drops at the final time step.
It can be seen that many small droplets are formed from narrow sections that extend radially from the collision center.
These small drops, however, form only on the sides, whereas the volume sections at the bottom and top contribute to relatively large drops.

In Figure~\ref{fig:dropcoll-cross-sec}, a section of the right drop from Figure~\ref{fig:teaser} has been cut out to reveal the inner structure of the volume contributions.
Interestingly, almost all droplets that develop after the collision are formed from volumes that are close to the collision center and propagate outward to the back side.
This structure bears some similarity to cracks in a solid material that propagate from the impact point.
In Figure~\ref{fig:dropcoll-s-surfaces}, separation surfaces $S$ are shown. Here, the breakup of inner disk and outer ring is apparent in the form of cone like structures in both droplets.
The surfaces within the cones have similar colors, indicating simultaneous separation of multiple droplets.

For this dataset we have additionally applied our method in reverse time direction, i.e., we computed the correspondence of the last simulation time step to the first one to reveal how exactly the initial two droplets contributed to the later smaller droplets.
The result is shown in Figure~\ref{fig:dropcoll-reverse}.
In Figure~\ref{fig:dropcoll-reverse1}, the two droplets at time $t_F=0.478\,\text{ms}$ are visualized by the PLIC interface.
They are colored by the connected component labels, and their contributions in the droplets at time $t_0=2.85\,\text{ms}$ are visualized in Figure~\ref{fig:dropcoll-reverse2}.
As can be seen, the volume from the red drop dominates on the upper right part of the droplets at later physical time.
Figure~\ref{fig:dropcoll-reverse3} shows a selected droplet that rotates after separation from the disk-shaped drop.
The boundary from the blue droplet is transparent to reveal the inner structure of the drop.
The interface between the two boundaries $B$ forms an S-shape, and there is distinguishable symmetry of the two volumes.
The investigation of the interplay between volume distribution and rotational motion of the feature could shed light on the details of drop dynamics, this is, however, out of scope of this paper.

\subsection{Oil Inclusions}
\label{sec:oil}

\begin{figure*}
  \centering
  \hfill%
  \subfigure[$0\,\textrm{ms}$]{\includegraphics[width=0.1\linewidth]{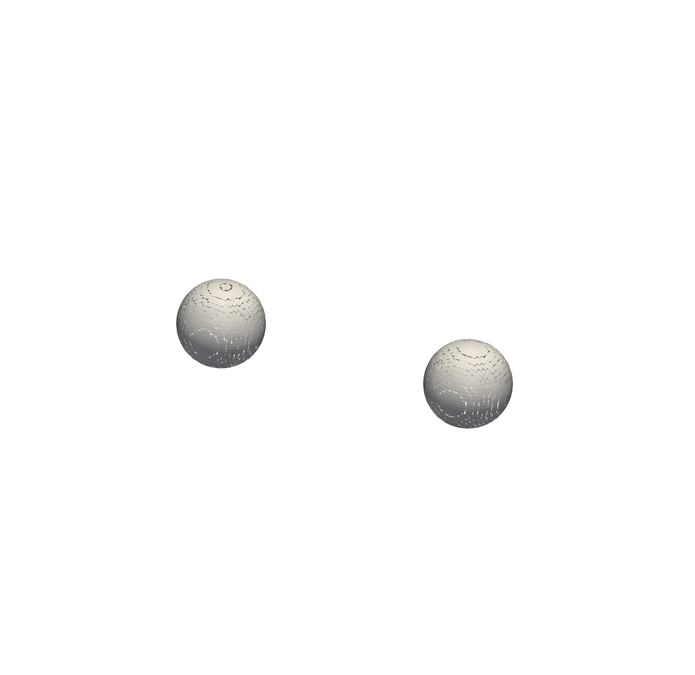}} 
  \hfill%
  \subfigure[$0.125\,\textrm{ms}$]{\includegraphics[width=0.1\linewidth]{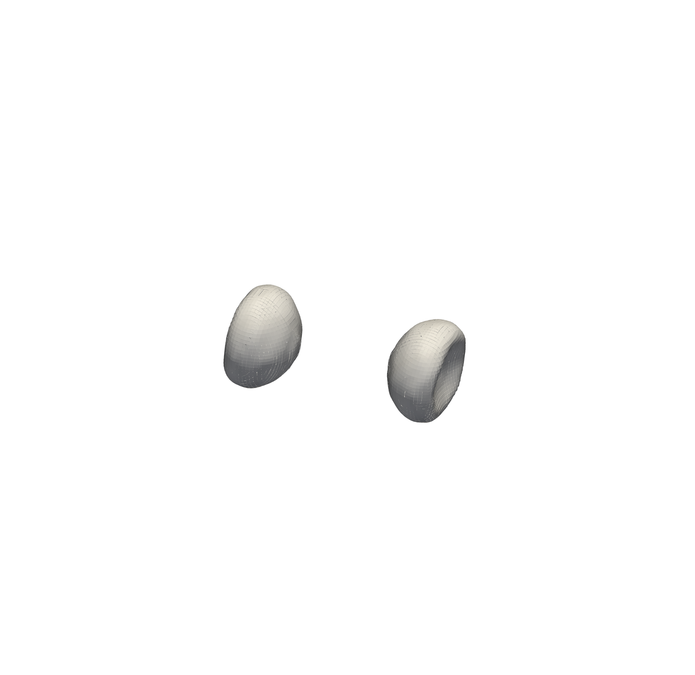}} 
  \hfill%
  \subfigure[$0.625\,\textrm{ms}$]{\includegraphics[width=0.1\linewidth]{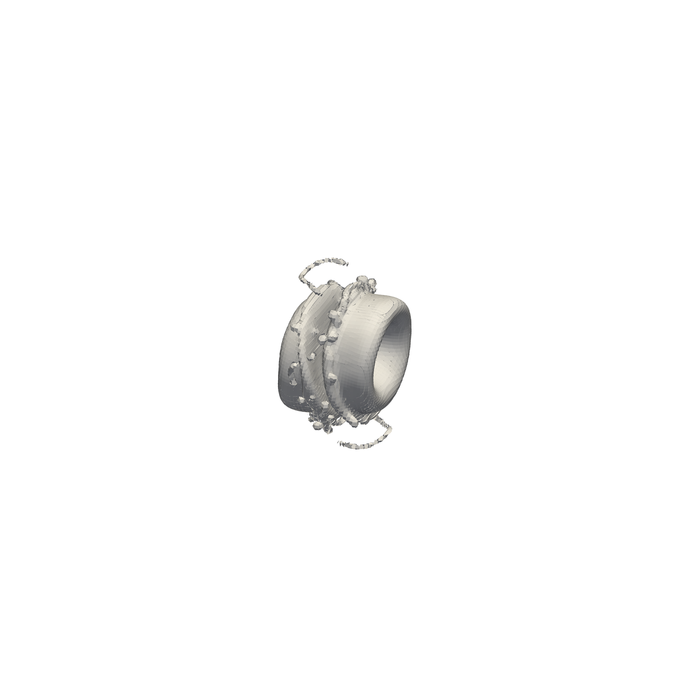}} 
  \hfill%
  \subfigure[$1.26\,\textrm{ms}$]{\includegraphics[width=0.1\linewidth]{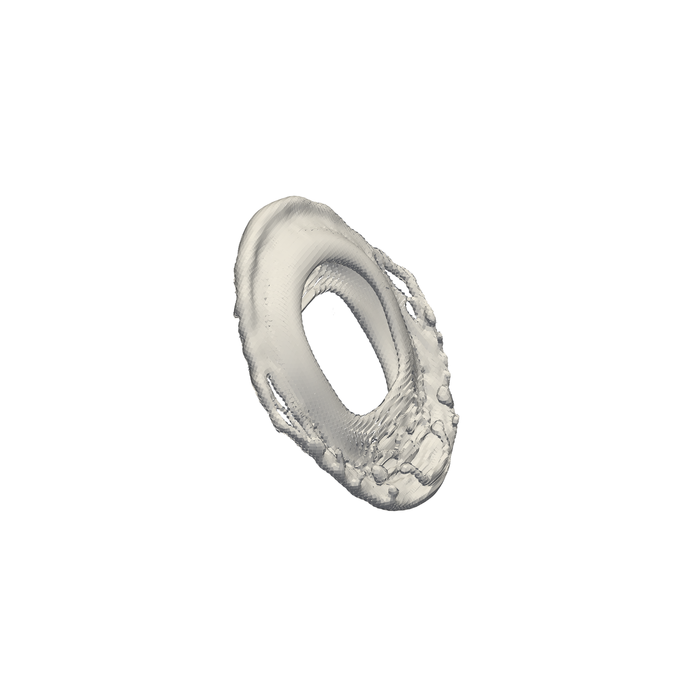}} 
  \hfill%
  \subfigure[$2.51\,\textrm{ms}$]{\includegraphics[width=0.1\linewidth]{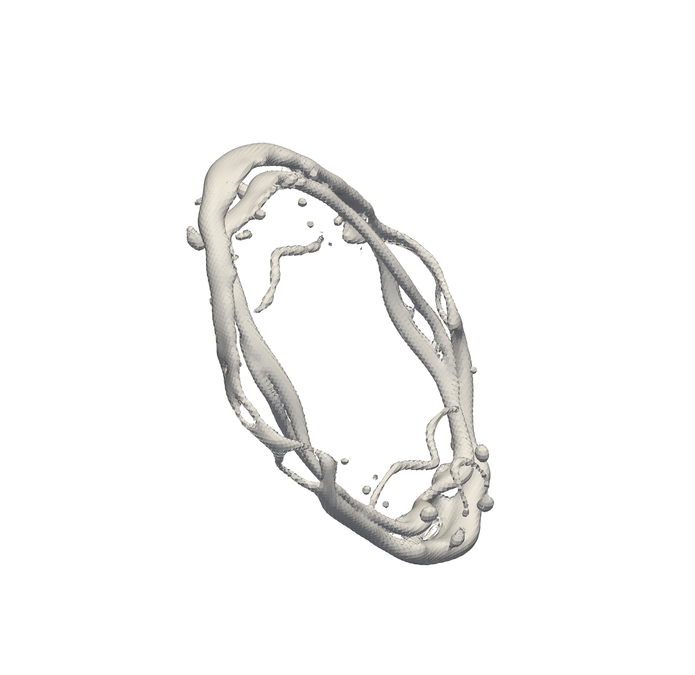}} 
  \hfill%
  \subfigure[$5.0\,\textrm{ms}$]{\includegraphics[width=0.1\linewidth]{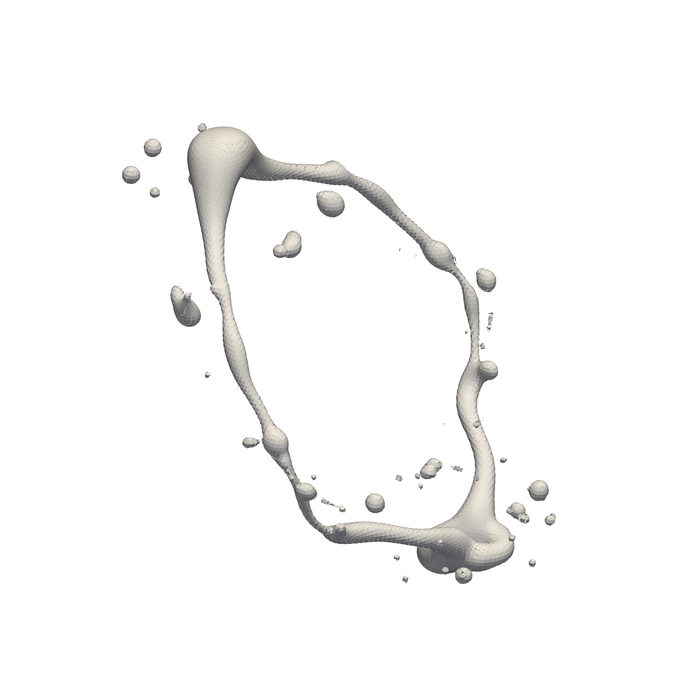}} 
  \hfill%
  \subfigure[$7.52\,\textrm{ms}$]{\includegraphics[width=0.1\linewidth]{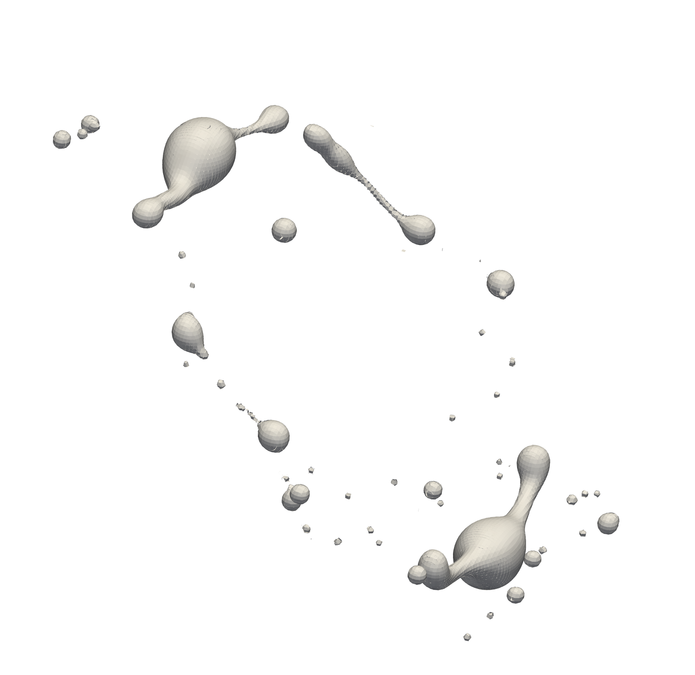}} 
  \hfill%
  \subfigure[$10\,\textrm{ms}$]{\includegraphics[width=0.1\linewidth]{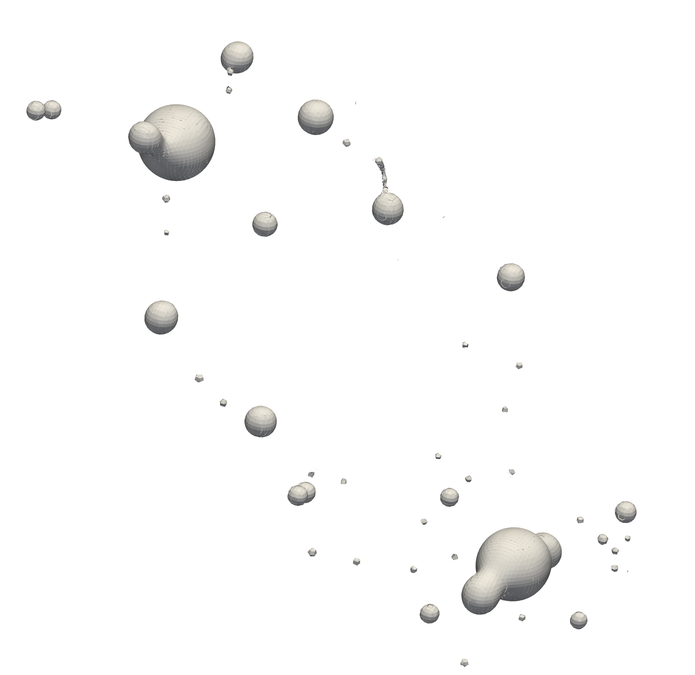}} 
  \hfill%
  \subfigure[$12.5\,\textrm{ms}$]{\includegraphics[width=0.1\linewidth]{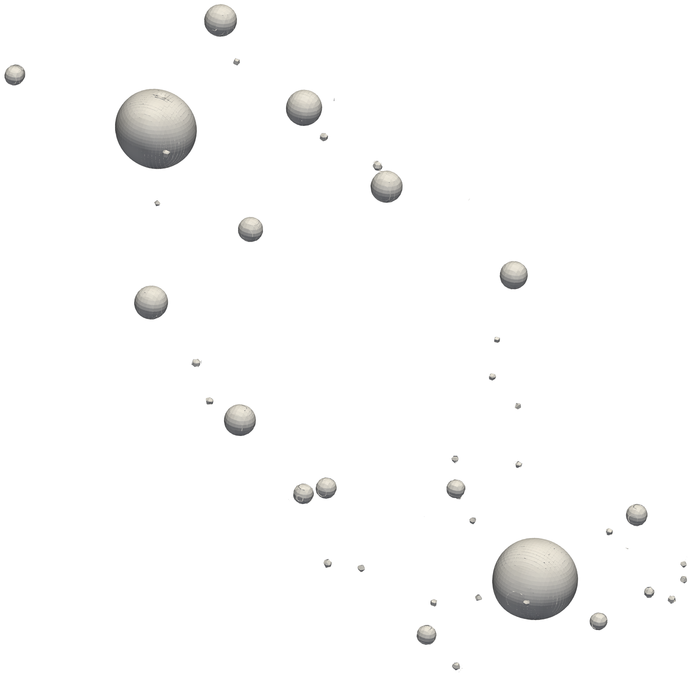}}
  \hfill%
  \caption{
    Extracted liquid interface for the simulation of \oil/.
    Even before collision, the inclusions deform strongly due to the high viscosity of the surrounding phase that dampens the initial momentum due to friction.
    After the collision, the ring disintegrates and the resulting inclusions form into spherical shapes.
  }\label{fig:oil}
\end{figure*}

The next dataset is a simulation of colliding oil inclusions in water surrounding.
For this dataset, $f$ represents volume fraction of oil.
The simulation domain consists of $256^3$ cells covering $1\,\text{cm}^3$, and $501$ time steps that span a time interval of $12.5\,\textrm{ms}$.
In Figure~\ref{fig:oil}, selected simulation time steps are shown.
The deformation of the oil inclusions differs strongly from
the water inclusions in the first dataset due to the presence of liquid surrounding.
Even before the collision, the initially spherical inclusions deform strongly and 
they lose a considerable amount of momentum due to friction.
In the final time steps after the collision, the inclusions quickly form spherical shapes and 
their velocity drops to almost zero.

\begin{figure}
  \centering
  \subfigure[]{\includegraphics[width=0.49\linewidth]{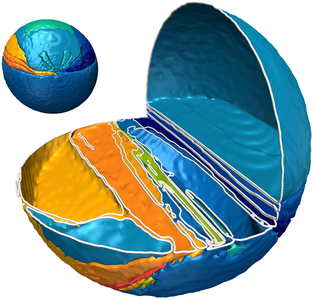}
    \label{fig:oil-cutout-labels}}
  \hfill%
  \hspace{-2mm}%
  \subfigure[]{\begin{overpic}[width=0.49\linewidth]{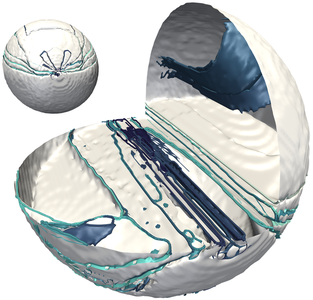}
      \linethickness{0.75pt}
      \multiput(50,58)(4,0){10}{\color{red}\line(1,0){2}}
      \multiput(50,92)(4,0){10}{\color{red}\line(1,0){2}}
      \multiput(50,58)(0,4){9}{\color{red}\line(0,1){2}}
      \multiput(90,58)(0,4){9}{\color{red}\line(0,1){2}}
    \end{overpic}
    \label{fig:oil-cutout-ssurfaces}}
  \subfigure{\includegraphics[width=0.6\linewidth]{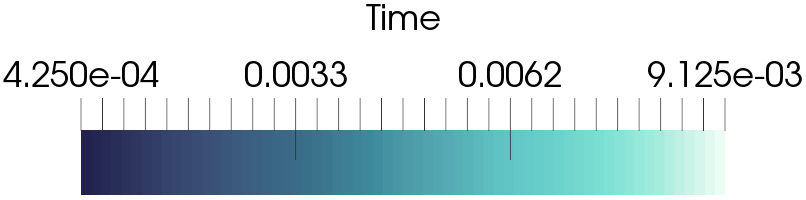} 
\label{fig:oil-cutout-legend}}
  \caption{
    Visualization of feature separation in \oil/ dataset. \subref{fig:oil-cutout-labels}~Separation boundaries $B$ reveal considerably different segmentation of feature than in the case of liquid-gas configuration. \subref{fig:oil-cutout-ssurfaces}~Temporal separation surfaces $S$ with white boundaries $B$ for context. Interestingly, the surface in the red box indicates separation, but lack of corresponding boundary suggests later merge. This is further investigated in Figure~\ref{fig:oil-merge}. Color legend for temporal surfaces.
  }\label{fig:oil-cutout}
\end{figure}

Figure~\ref{fig:oil-cutout-labels} shows the boundaries $B_{t_0}^{t_F}$ with the upper left part of the front drop cut out to reveal the inner segmentation.
The structure is quite distinct from the one in the previous dataset, as it is less ordered and contains many small segments in the central part.

\begin{figure*}
  \centering
  \subfigure[]{\begin{overpic}[width=0.63\linewidth]{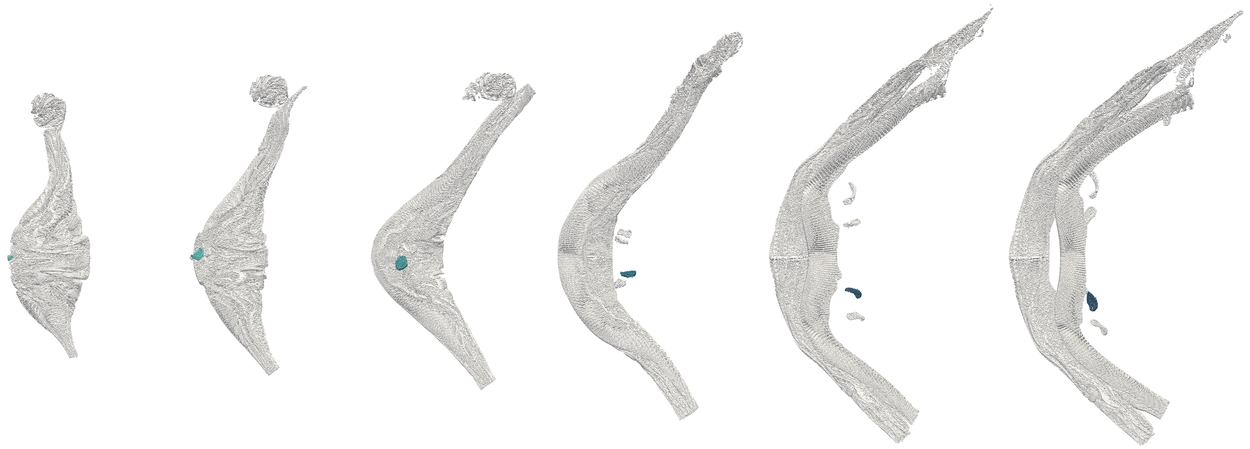}
      \linethickness{0.75pt}
      \multiput(67,11)(2,0){2}{\color{red}\line(1,0){1}}
      \multiput(67,11)(0,2){2}{\color{red}\line(0,1){1}}
      \multiput(67,15)(2,0){2}{\color{red}\line(1,0){1}}
      \multiput(71,11)(0,2){2}{\color{red}\line(0,1){1}}
      \put(71,15){\small $(i)$}

      \multiput(30.5,13)(2,0){2}{\color{red}\line(1,0){1}}
      \multiput(30.5,13)(0,2){2}{\color{red}\line(0,1){1}}
      \multiput(30.5,17)(2,0){2}{\color{red}\line(1,0){1}}
      \multiput(34.5,13)(0,2){2}{\color{red}\line(0,1){1}}
      \put(34.5,17){\small $(ii)$}

      \put(17,1.5){\small Time}
      \put(30,1){\color{black}\vector(-1,0){20}}
    \end{overpic} 
    \label{fig:oil-label14-tracking}}
  \hfill%
  \subfigure[]{\begin{overpic}[width=0.32\linewidth]{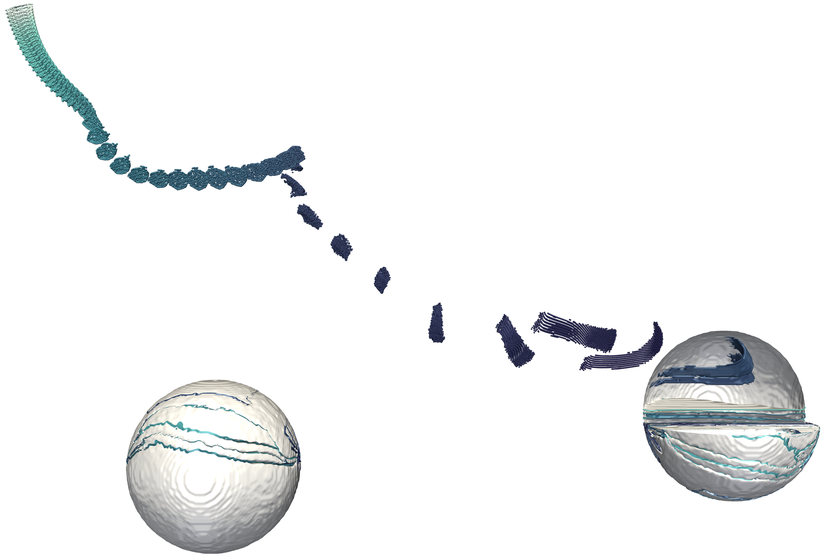}
      \linethickness{0.75pt}
      \multiput(32,45)(4,0){2}{\color{red}\line(1,0){2}}
      \multiput(32,53)(4,0){2}{\color{red}\line(1,0){2}}
      \multiput(32,45)(0,4){2}{\color{red}\line(0,1){2}}
      \multiput(40,45)(0,4){2}{\color{red}\line(0,1){2}}
      \put(40,53){\small $(i)$}

      \multiput(10,45)(4,0){2}{\color{red}\line(1,0){2}}
      \multiput(10,53)(4,0){2}{\color{red}\line(1,0){2}}
      \multiput(10,45)(0,4){2}{\color{red}\line(0,1){2}}
      \multiput(18,45)(0,4){2}{\color{red}\line(0,1){2}}
      \put(18,53){\small $(ii)$}
    \end{overpic} 
    \label{fig:oil-label14}}
  \caption{
Investigation of the segment from Figure~\ref{fig:oil-cutout-ssurfaces}, with time color-coded on the segment of interest. 
\subref{fig:oil-label14-tracking}~white particles correspond to the whole volume enclosed be the separation boundary. 
The small inclusion separates from the whole segment  (box $(i)$), and merges with it again (box $(ii)$).
\subref{fig:oil-label14}~Side view of the inclusions at initial time step with particles shown for intermediate time steps (i.e., every $5$th simulation step). Changes of movement direction correspond to the split and merge (red boxes). Color legend same as in Figure~\ref{fig:oil-cutout}.
  }\label{fig:oil-merge}
\end{figure*}

The separation surfaces for the same configuration are shown in Figure~\ref{fig:oil-cutout-ssurfaces}.
Interestingly, in the upper segment, there is a closed surface (marked with red box) that indicates split followed by merge event.
With our technique, we can further investigate this part by selecting the particles comprising the enclosing segment (and therefore, a developing inclusion), as visualized in Figure~\ref{fig:oil-label14-tracking}. Here, the first time step was selected just before the split, the second shows the small part that separated (red box) and the last three show the re-merging process. 
In Figure~\ref{fig:oil-label14} we identified the particles that belong to the splitting part and shown them for all stored time steps (i.e., every $5$th simulation step), with time color-coded. 
The bend parts clearly indicate the split and merge (red boxes).

This dataset demonstrates the ability of our technique to perform a detailed inspection of the separation processes.
%
\subsection{Non-Newtonian Jet}
\label{sec:results-jet}

\begin{figure}[t]
  \centering
  \includegraphics[width=0.9\linewidth]{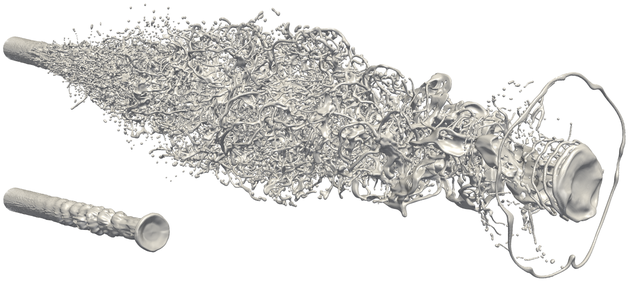}
  \caption{
    Selected time steps in the \textit{Jet} dataset.
    On the bottom, the jet at time step $t_0=1.6\,\text{ms}$ is about to break up into a multitude of droplets and ligaments.
    On the top, a moderately developed jet at $t_F=3.97\,\text{ms}$ is shown.
  }\label{fig:jet-overview}
\end{figure}
\begin{figure}[t]
  \centering
  \includegraphics[width=0.99\linewidth]{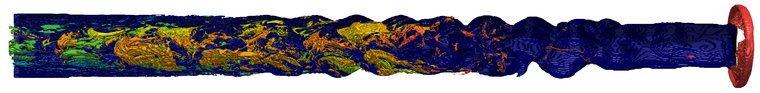}
  \includegraphics[width=0.99\linewidth]{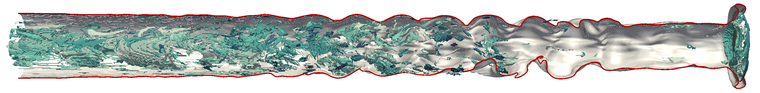}
  \includegraphics[width=0.6\linewidth]{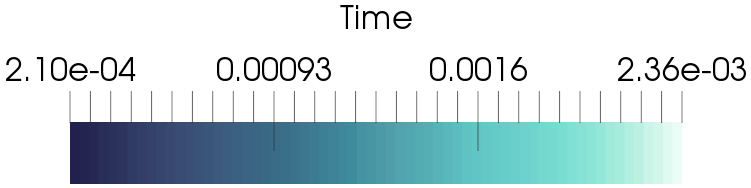} 
  \caption{
Separation boundaries $B$ (top) and temporal separation surfaces $S$ (bottom) in the \jet/ dataset, with front part cut out to reveal inner structure. The jet disintegrates mostly in the rear part, and at the very front which bends back. Small dark blue separation surfaces indicate that small droplets disintegrate early.
  }\label{fig:jet-ss-boundaries}
\end{figure}
The last dataset is a simulation of the injection of a jet made up of a non-Newtonian shear thinning aqueous solution of Praestol 2500 (0.3\% weight) into air at $p_{\text{atm}} = 1\,\text{bar}$.
 The jet is introduced into a rectangular computational domain with a diameter of $\textrm{D} = 1.2\,\text{cm}$ and with the velocity profile from a short nozzle with a mean velocity of $u = 75\,\text{m/s}$, and therefore a Reynolds number $Re=19\,000$.
The domain has a size of $42\,\textrm{D}$ in the direction of the injection and of $10\,\textrm{D}$ in each of the directions of the injection plane.
The domain is discretized over $2688 \times 512 \times 512$ cells.
The temporal development of the jet is sketched in Figure~\ref{fig:jet-overview}.
At $t_0=1.60\,\textrm{ms}$, the jet has just entered the computational domain and has started to interact with the surrounding gas.
The resulting effects include the bending back of the jet tip and the development of surface waves on the jet core.
Due to the high Reynolds number, the jet becomes unstable very quickly, and at $t_F=3.97\,\textrm{ms}$, the core has expanded substantially in radial direction and has mostly broken up into ligaments and droplets---the atomization has begun.

Figure~\ref{fig:jet-ss-boundaries} shows the boundaries $B$ and separation surfaces $S$ at $t_0$. Small polygons 
have been removed for clear overview. Apparently, in the investigated time interval, the front of the jet does not undergo strong disintegration (except for the tip), whereas in the rear, many segments imply the origins of elongated droplets that develop later. The separation surfaces at the bottom of the figure are uniformly distributed, although small droplets detach earlier, as indicated by small dark blue surfaces.

\begin{figure*}[t]
  \centering
  \subfigure[]{\includegraphics[width=0.7\linewidth]{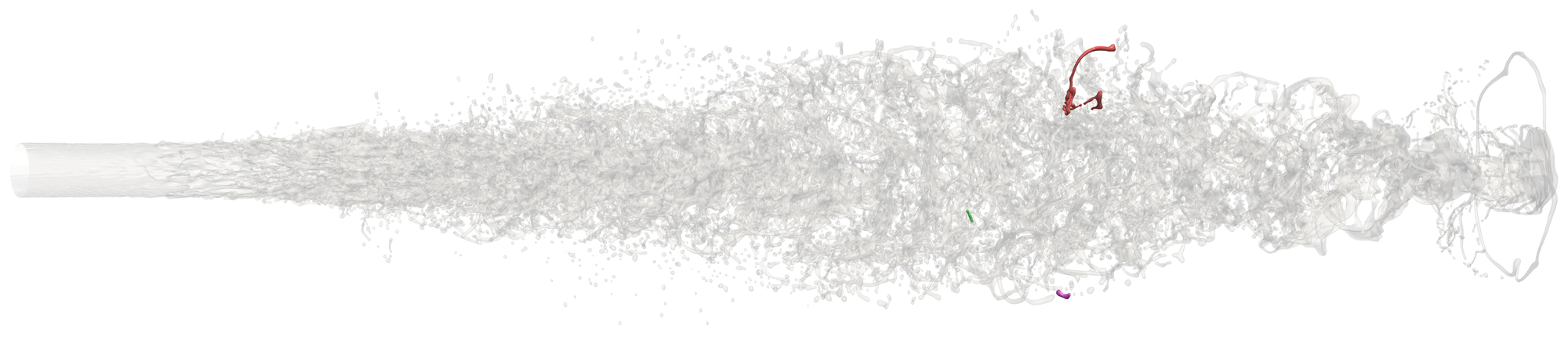}\label{fig:jet-a}}
  \subfigure[]{\includegraphics[width=0.7\linewidth]{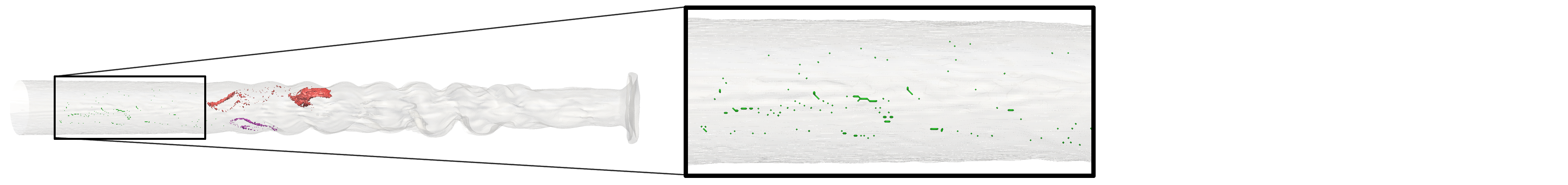}\label{fig:jet-b}}
  \caption{
    Visualization of feature separation for the \textit{Jet} dataset.
    \subref{fig:jet-a} Selected ligaments (colored by label) at time $t_F$ are shown in the context of the whole jet (transparent gray).
    \subref{fig:jet-b} The visualization of the boundaries $B$ at time $t_0$ reveals the intricate, elongated structures of the corresponding volumetric contributions. 
  }\label{fig:jet-init}
\end{figure*}

To better understand the atomization process, we use the proposed visualization method to investigate the origin of ligaments within the liquid core.
In Figure~\ref{fig:jet-b}, we can observe the spatial origin at $t_0$ for a selected number of ligaments, and their position at $t_0 +\Delta t$ is displayed in Figure~\ref{fig:jet-a}.
In Figure~\ref{fig:jet-a}, the three features are in close proximity in stream-wise $x$-direction.
This is surprising, as their original liquid mass in Figure~\ref{fig:jet-b} was distributed over a rather large range in $x$-direction.
In Figure~\ref{fig:jet-b}, the liquid mass of the ligaments is scattered within the jet core.
This is especially noticeable for the droplet colored in green, which can be seen in the zoomed cutout.
This indicates a strong influence of the flow field inside the jet core, as the liquid mass has to combine before separating from the core.

In Figure~\ref{fig:jet-particles}, the temporal evolution of one selected feature (colored) is shown with the entire jet (gray) in the background.
The temporal evolution is from $t_0$ at the top to $t_0 +\Delta t$ at the bottom.
The development in the first three time steps confirms our assumption from Figure~\ref{fig:jet-a}.
The liquid mass, which will later form the ligament, is initially scattered over one third of the jet length.
With time, the liquid mass near the jet surface is being decelerated due to the interaction with the surrounding atmosphere.
 This allows the rest of the mass, which started out more central in the jet, to catch up, and by the third time step, the bulk of the liquid mass is located within the first surface wave, while a few parts are located in the next wave.
From the underlying jet, we can observe that the waves are already starting to deform strongly from the shape of a periodic wave.
At the next time step, nearly all of the liquid mass has merged and moved even closer toward the surface.
The liquid mass has reached its most compact form.
In the next time step, the surface wave has disintegrated and the liquid mass has become a complex branching ligament.
With time, the ligament moves further in radial direction, interacting more strongly with the atmosphere and therefore is stretched in $x$-direction, until we can observe the first droplet breaking off the ligament in the final time step.

With the proposed visualization method, we were able to make interesting observations.
Contrary to the impression we get from Figure~\ref{fig:jet-overview} that the jet is mainly expanding in radial direction, we observed in Figure~\ref{fig:jet-a} that features which are in close proximity in $x$-direction at $t_0 +\Delta t$ originate from rather distant positions at $t_0$.
From this, we can infer that the complex nature of the velocity field leads to features moving at higher velocity in $x$-direction.
We could also observe that this is even true for the velocity field within the jet core.
Contrary to the rather solid seeming core in Figure~\ref{fig:jet-overview}, both Figure~\ref{fig:jet-a} and especially Figure~\ref{fig:jet-particles} present how the flow field inside the core is creating the analyzed features from liquid mass scattered throughout the core.
Furthermore, Figure~\ref{fig:jet-particles} is an excellent representation of the different steps of primary jet break-up showing the formation of surface waves, the deformation of the waves, the rupturing of the waves into ligaments, and the stretching and break-up of the ligaments into droplets.

The presented method allows for a simple and intuitive way to analyze the spatial and temporal development of the primary break-up of liquid jets.
Thanks to this method an insight into the movement patters of select features of the break-up process could be gained which would otherwise have been hard and unlikely to obtain.

\def\jetwidth{0.7}
\begin{figure*}[t]
  \centering
  \includegraphics[width=\jetwidth\linewidth]{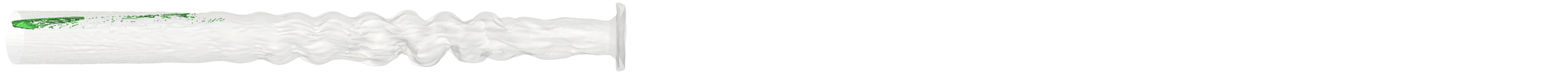}
  \includegraphics[width=\jetwidth\linewidth]{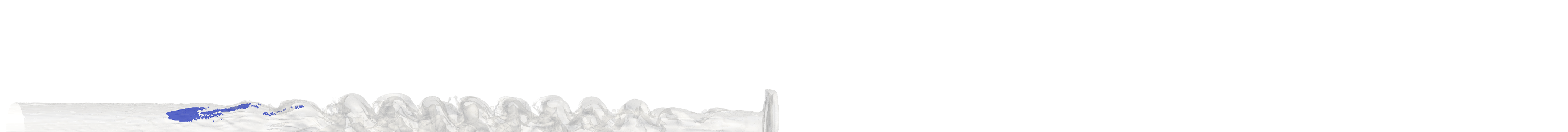}
  \includegraphics[width=\jetwidth\linewidth]{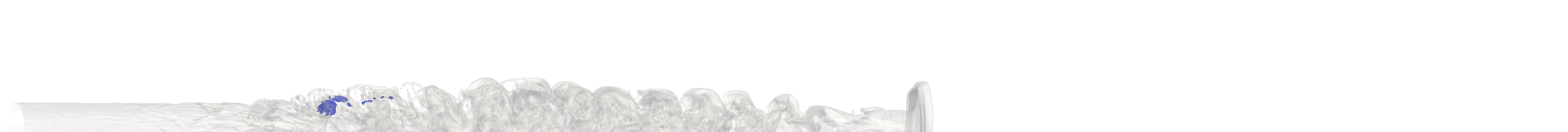}
  \includegraphics[width=\jetwidth\linewidth]{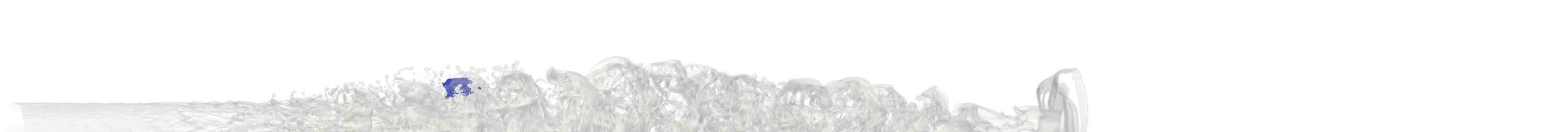}
  \includegraphics[width=\jetwidth\linewidth]{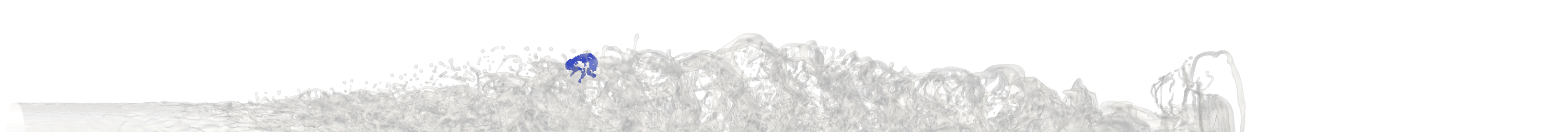}
  \includegraphics[width=\jetwidth\linewidth]{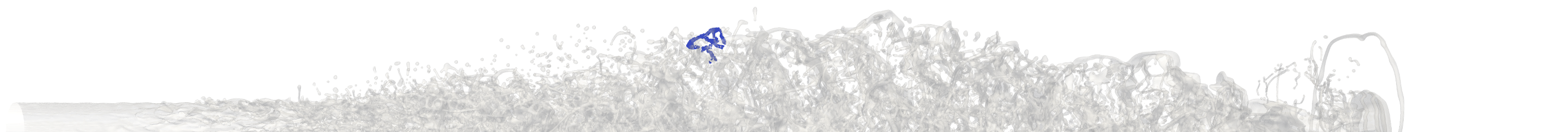}
  \includegraphics[width=\jetwidth\linewidth]{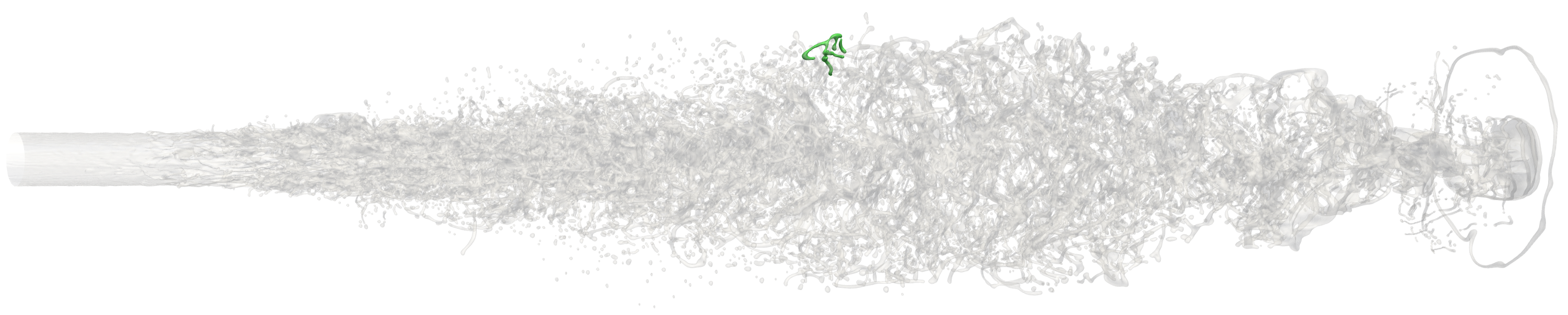}
  \caption{
    Selected volumetric contributions in the \jet/ dataset at $t_0$ (top) and the corresponding ligament at $t_F$ (bottom).
    In the middle figures, the temporal evolution of particles for the selected contribution is shown at intermediate time steps.
    The volumetric contribution curls into the final ligament as it is pushed away from the jet core by the surrounding air.
  }\label{fig:jet-particles}
\end{figure*}

\subsection{Performance}
\label{sec:performance}

The computation of the visualization for the \oil/ and
\dropcoll/ datasets was performed on a commodity desktop with Intel i7 3.6 GHz processor (4 cores) and $32\,\textrm{GiB}$ RAM.
For the \jet/ dataset, we have used $64$ nodes on a Cray XC40 System, each with two Intel Xeon E5 processors and $128\,\textrm{GiB}$ of memory, and one process running per node.
Table~\ref{tab:performance} shows the configurations (i.e., number of simulation time steps and the number of seeded particles) and timings (i.e., total time, average time for one advection step and extraction of $S$, and time for extraction of $B$) for the three datasets.
Although all datasets have comparable numbers of particles, the computation of the \jet/ dataset is not much faster, despite the parallelization. This is due to the fact that with the data parallelism in the multi-process configuration, the particles are not evenly distributed among the processes.
Moreover, the exchange of particles across the subdomains introduces some overhead after each integration step.

%
\begin{table}[b]
\caption{Computation times for the three datasets.
For \dropcoll/ (Drops) and \oil/ (Oil), the measurement was performed on a desktop computer, and on the Cray system for the \jet/ (Jet) dataset.
``Time'' is the total computation time, ``Adv. Step+$S$'' is the time for one step (done for each simulation step) of  advection and extraction of $S$. ``$B$'' is time for extraction of boundaries (done once).\label{tab:performance}}
  \centering
  \begin{tabular}{@{}l c D{.}{.}{2.1} D{.}{.}{2.1} D{.}{.}{2.1} D{.}{.}{2.1} }
  \toprule
    Dataset & \#Steps & \multicolumn{1}{c}{\#Particles} & \multicolumn{1}{c}{Adv. Step+$S$} & \multicolumn{1}{c}{$B$} & \multicolumn{1}{c}{Total} \\
 	\midrule
    Drops & $400$ & 1.1\textrm{e}6 & 54.7\,\textrm{s} & 5.8\,\textrm{s} & 6.8\,\textrm{h} \\
    Oil & $500$ & 1.1\textrm{e}6 & 54.8\,\textrm{s} & 6.4\,\textrm{s} &  8.3\,\textrm{h} \\ 
    Jet & $237$ & 1.7\textrm{e}6 & 29.5\,\textrm{s} & 6.1\,\textrm{s} & 3.9\,\textrm{h} \\
    \bottomrule
  \end{tabular}
\end{table}

We have also investigated the performance dependency on parameter $r$. As Table~\ref{tab:parameter-r} shows, in case of \dropcoll/, the difference between $r=0$ and $r=1$ is relatively small arguably due to better usage of thread-level parallelism. 
For $r=2$, on the other hand, the number of particles increases the computation time substantially.
For the \jet/, 
it can be seen that with $8$-fold increase in particle number, there is $8$-fold increase in time for advection step and  the total time increases about $5$ times, suggesting that the interprocess communication constitutes considerable time which is less dependent on the number of particles.
\begin{table}[b]
\caption{Computation times for different value of parameter $r$ for \dropcoll/ and \jet/ datasets. Same notation as in Table~\ref{tab:performance}.\label{tab:parameter-r}}
  \centering
  \begin{tabular}{@{}l c D{.}{.}{2.1} D{.}{.}{2.1} D{.}{.}{2.1} D{.}{.}{2.1}}
  \toprule
    Dataset & $r$ & \multicolumn{1}{c}{\#Particles} & \multicolumn{1}{c}{Adv. Step+$S$} &  \multicolumn{1}{c}{$B$} &  \multicolumn{1}{c}{Total} \\
    \midrule
    Drops & $0$ & 1.8\textrm{e}4 &  2.1\,\textrm{s} & 0.3\,\textrm{s} & 0.9\,\textrm{h} \\ 
    Drops & $1$ & 1.4\textrm{e}5 &  7.9\,\textrm{s} & 0.9\,\textrm{s} & 1.5\,\textrm{h} \\ 
    Drops & $2$ & 1.1\textrm{e}6 & 54.7\,\textrm{s} & 5.8\,\textrm{s} & 6.8\,\textrm{h} \\ 
    Jet & $0$ & 1.7\textrm{e}6 & 29.5\,\textrm{s} & 6.1\,\textrm{s} & 3.9\,\textrm{h} \\
    Jet & $1$ & 13.6\textrm{e}6 & 234.0\,\textrm{s} & 46.2\,\textrm{s} & 18.9\,\textrm{h} \\
    \bottomrule
  \end{tabular}
\end{table}

For the visualization we have used a commodity desktop for all datasets, since the data produced by our technique (i.e., the particle data and mesh for boundaries $B$) does not exceed $1\,\textrm{GiB}$, even for the \jet/ dataset.


\subsection{Analysis of the Corrector in Multiphase Flow}
\label{sec:error}
\begin{figure}[t]
  \centering
  \subfigure[]{%
    \includegraphics[width=0.95\linewidth]{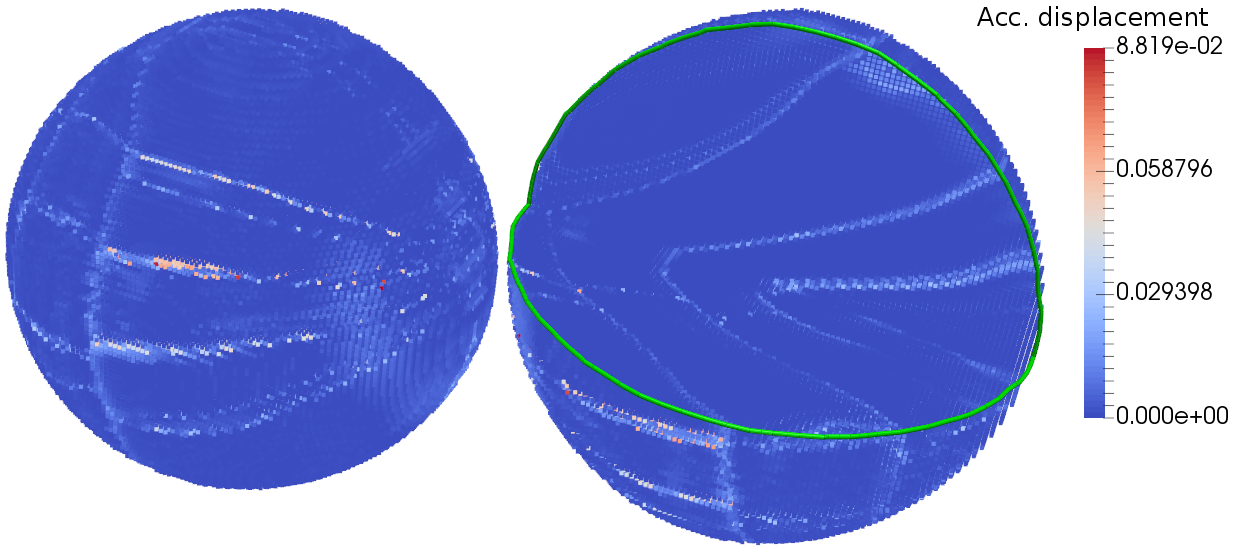} 
    \label{fig:acc-corrector1}}
  \subfigure[]{%
    \includegraphics[width=0.98\linewidth,clip=true,trim=8cm 0px 0px 0px]{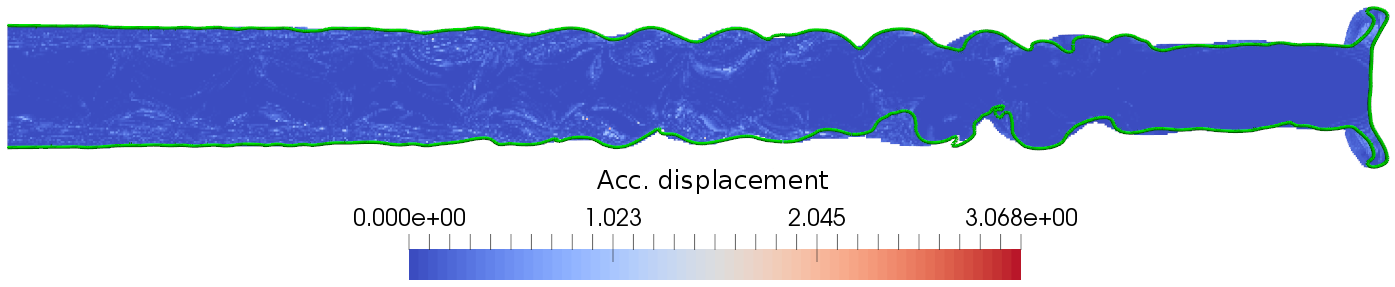}
    \label{fig:acc-corrector2}}
  \caption{
    Visualization of the accumulated displacement for the investigated multiphase flow datasets.
    Cross-sections are shown as green lines.
    The displacement values $\epsilon_\particle$ of the integrated particles are mapped on the corresponding seed points.
    Both in the \dropcoll/~\subref{fig:acc-corrector1} and \jet/~\subref{fig:acc-corrector2} datasets, larger values correspond to the boundary regions, which are most unstable.
  }\label{fig:acc-corrector}
\end{figure}
To quantify the amount of applied correction that is necessary to keep the particles within the initial phase in multiphase flow (see Section~\ref{sec:vof-corrector}), we accumulate the introduced displacement for each particle over all time steps $i$:
\begin{equation}
\epsilon_{\particle}=\sum_{i\in I} ( |\vecx'_{\particle,i}-\vecx_{\particle,i}|+|\vecx''_{\particle,i}-\vecx'_{\particle,i}|+|\vecx'''_{\particle,i}-\vecx''_{\particle,i}|) \, ,
\end{equation}
where the first term is the translation with the displacement vector of the nearest phase-consistent particle (Figure~\ref{fig:smart-corrector}), the second term in the sum corresponds to the translation to the nearest cell with $f>0$ (Figure~\ref{fig:vof-corrector}), and the third term is the translation to the PLIC patch in the interface cells (Figure~\ref{fig:plic-corrector}).
For the analysis, we visualize the value $\epsilon_\particle$ at the seed positions $\vec{s}_\particle$.
In Figures~\ref{fig:acc-corrector1} and~\ref{fig:acc-corrector2} the displacement $\epsilon_\particle$ is shown for the \dropcoll/ and \jet/ datasets, respectively.
The particles that were translated most correspond with the boundary positions.
This is easy to interpret, since the particles that most probably leave the liquid phase are close to the feature interface, i.e., at unstable regions.
Although the maximum value of the accumulated correction is relatively large ($0.07$ of the domain size for \dropcoll/ and $0.06$ of the domain size for the \jet/), it occurs only for small number of particles, and the correction for the majority of particles is considerably smaller (as indicated by pale blue color in the figures).

\subsubsection*{Robustness}
\label{sec:robustness}
\def\trimwidth{18.75}
\begin{figure}[t]
  \centering
  \subfigure[$50$ steps; no correction.]{%
    \includegraphics[width=0.48\linewidth,clip=true,trim=0px \trimwidth mm 0px 0mm]{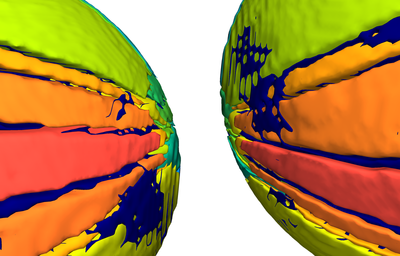}
    \label{fig:robustness-no-corr-e8}}
  \hfill%
  \subfigure[$400$ steps; no correction.]{%
    \includegraphics[width=0.48\linewidth,clip=true,trim=0px \trimwidth mm 0px 0mm]{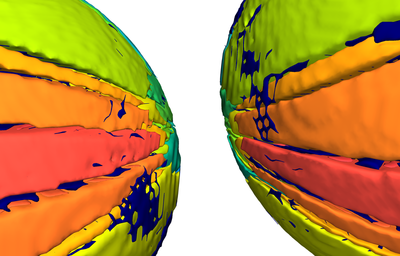}
    \label{fig:robustness-no-corr-e1}}
  \subfigure[$50$ steps; no displ. vec.]{%
    \begin{overpic}[width=0.48\linewidth,clip=true,trim=0px \trimwidth mm 0px 0mm]{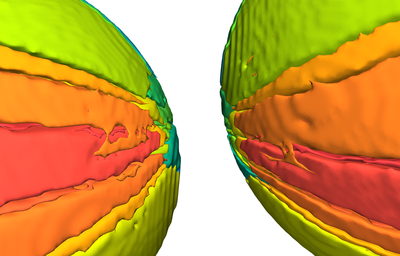}
      \linethickness{1pt}
      \put(20,21){\color{black}\line(1,0){20}}
      \put(20,9){\color{black}\line(1,0){20}}
      \put(20,9){\color{black}\line(0,1){12}}
      \put(40,9){\color{black}\line(0,1){12}}
    \end{overpic}
    \label{fig:robustness-no-smart-e8}}
  \hfill%
  \subfigure[$400$ steps; no displ. vec.]{%
    \includegraphics[width=0.48\linewidth,clip=true,trim=0px \trimwidth mm 0px 0mm]{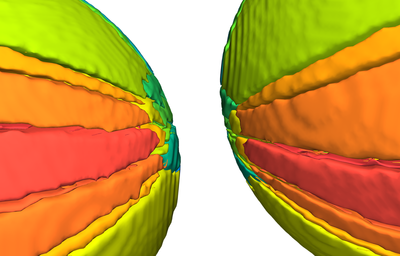}
    \label{fig:robustness-no-smart-e1}}
  \subfigure[$50$ steps; full correction.]{%
    \includegraphics[width=0.48\linewidth,clip=true,trim=0px \trimwidth mm 0px 0mm]{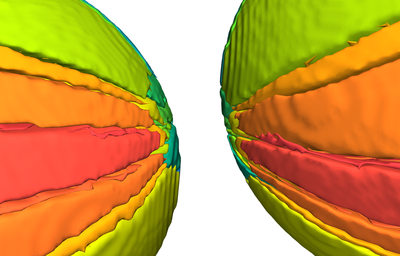}
    \label{fig:robustness-smart-e8}}
  \hfill%
  \subfigure[$400$ steps; full correction.]{%
    \includegraphics[width=0.48\linewidth,clip=true,trim=0px \trimwidth mm 0px 0mm]{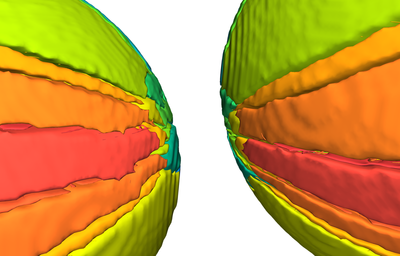}
    \label{fig:robustness-smart-e1}}
  \caption{
Analysis of corrector robustness for $50$ (left column) and $400$ (right column) time steps. \subref{fig:robustness-no-corr-e8},~\subref{fig:robustness-no-corr-e1}~No correction applied, with many stray particles (corresponding blue surfaces). 
\subref{fig:robustness-no-smart-e8},~\subref{fig:robustness-no-smart-e1} Only second and third stage. 
\subref{fig:robustness-smart-e8},~\subref{fig:robustness-smart-e1} full correction. 
Incorrectly assigned particles at the collision region in~\subref{fig:robustness-no-smart-e8} are corrected either with better temporal resolution \subref{fig:robustness-no-smart-e1} or with full correction \subref{fig:robustness-smart-e8}.
  }\label{fig:robustness}
\end{figure}
In Figure~\ref{fig:robustness} we also investigate the sensitivity of the particle advection to temporal resolution of data and different variants of the particle corrector. The selected droplet region from the \dropcoll/ dataset is particularly prone to error due to the collision which is difficult to capture with integration. For the analysis, we use the temporal resolution of $50$ (left column) and $400$ (right column) simulation time steps spanning the same time interval. The top row shows the resulting boundary reconstruction for particles with no correction applied, in the middle row, the particles were moved to the closest cell and consequently to the PLIC patch without the adjustment by the displacement vector. In the bottom row, full correction is applied. With no correction, even with high temporal resolution, there are regions containing particles that left the initial phase, as indicated by dark blue surfaces. Interestingly, for two-stage correction, the stray particles are gone, but in Figure~\ref{fig:robustness-no-smart-e8}, some particles at the front are incorrectly assigned. With full correction, little difference can be seen between both resolutions. This shows that our method can provide reliable results also for lower temporal resolution, although in our experiments we used higher temporal resolution to further increase reliability of the results. 

\section{Discussion}

As already mentioned in Sections~\ref{sec:simulation-data} and~\ref{sec:phase-consistent-traj}, there are two problems with tracking particles in multiphase flow. 
First, reduced temporal resolution of the simulation data does not allow for accurate integration due to the dynamics of phases. 
Particle advection is particularly challenging where phases collide, as their momentum changes rapidly. 
Second, the involved surface tension forces lead to very complex behavior, e.g., in regions where inclusions disintegrate.
Several methods have been proposed to find mesh and volume correspondences. 
However, those methods either work only on 2D manifolds, or provide correspondence based on optimization problems that 
typically cannot guarantee
physical correctness, e.g., by ignoring the underlying velocity field. 
For instance, the Wasserstein distances may result in physically inaccurate distant correspondences. 
Moreover, applied to volume, the correspondences from one cell usually scatter among many cells, which additionally complicates the choice of the correct particle path.

Therefore, we devised a new correction method which, admittedly, still has some limitations.
Due to the translation to the nearest cell, this method introduces asymmetry in the advection, i.e., the particles have different trajectories depending on the integration type (forward or backward in time).
In fact, we have observed that for the reverse advection in the \dropcoll/ dataset, the drop in the bottom right in Figure~\ref{fig:dropcoll-reverse2} is identified as coming from only the red drop in Figure~\ref{fig:dropcoll-reverse1}, whereas in the forward advection, we have observed that both drops contributed to it.
We have investigated several approaches to solve this problem, which, however, due to its complexity, remains beyond the scope of this work.
In the future, we would like to further pursue possible solutions to the asymmetry.

\section{Conclusion and Future Work}

In this paper, we have proposed a novel visualization method for the analysis of the separation of features. Whereas the boundaries $B$ determine feature volumes, the separation surfaces $S$ provide the information on the time point at which a given feature separation has occurred.
Hence, both methods complement each other to provide a comprehensive space-time visualization of feature separation.

For the analysis of multiphase flow, we have introduced a corrector method that utilizes the scalar field representing one of the fluid phases in order to ensure phase-consistent particle trajectories.
We have demonstrated the utility of our method for liquid-gas and liquid-liquid flow datasets. However, it can be as well applied to more complex multiphase simulations (e.g, with gas and two liquid components), since each phase is usually described by a separate scalar field, and therefore allows for direct application of our method.
Similarly, transitions between liquid and solid states could be investigated, where both states are stored as separate field data and therefore can be used to distinguish between the phases.
\bibliographystyle{abbrv}
\bibliography{KarchTopologyFeatureDynamics}
\end{document}